\documentclass[onecolumn,trackchanges]{aastex62}
\usepackage{graphicx}
\usepackage{amssymb}                    
\usepackage{ulem}
\usepackage{float}
\usepackage{rotating}
\usepackage{enumitem}
\usepackage{rotating}
\usepackage{makecell}
\usepackage{appendix}
\usepackage{fixmath}

\usepackage{color}

\usepackage{times}
\usepackage{amsmath}
\submitjournal{ApJ}

\shortauthors{Athiray et al.}
\shorttitle{High Temperature Slope}
\newcommand{\foxsi}{\textit{FOXSI-2}}
\newcommand{\aia}{\textit{SDO}/AIA}
\newcommand{\xrt}{\textit{Hinode}/XRT}
\newcommand{\eis}{\textit{Hinode}/EIS}
\newcommand{\magixs}{\textit{MaGIXS}}
\newcommand{\rhessi}{\textit{RHESSI}}
\newcommand{\twait}[1][]{t_{\textup{wait}#1}}
\begin{document}


\title{Solar Active Region Heating Diagnostics from High Temperature Emission using the Marshall Grazing Incidence X-ray Spectrometer (MaGIXS)}

\correspondingauthor{P.S. Athiray}
\email{athiray.panchap@nasa.gov}

\author[0000-0002-4454-147X]{P.S. Athiray}
\affil{NASA Marshall Space Flight Center, ST13, Huntsville, AL 35812}

\author[0000-0002-5608-531X]{Amy R.\ Winebarger}
\affil{NASA Marshall Space Flight Center, ST13, Huntsville, AL 35812}

\author[0000-0001-9642-6089]{Will T.\ Barnes}
\affil{Lockheed Martin Solar and Astrophysics Laboratory, Palo Alto, CA 94304}
\affil{Bay Area Environmental Research Institute, Moffett Field, CA 94952}

\author{Stephen J.\ Bradshaw}
\affil{Department of Physics \& Astronomy, Rice University, Houston, TX 77251}

\author{Sabrina Savage}
\affil{NASA Marshall Space Flight Center, ST13, Huntsville, AL 35812}

\author[0000-0001-6102-6851]{Harry P.\ Warren}
\affil{Space Science Division, Naval Research Laboratory, Washington, DC 20375 USA}

\author{Ken Kobayashi}
\affil{NASA Marshall Space Flight Center, ST13, Huntsville, AL 35812}

\author{Patrick Champey}
\affil{NASA Marshall Space Flight Center, ST13, Huntsville, AL 35812}

\author{Leon Golub}
\affil{Smithsonian Astrophysical Observatory, Cambridge, MA 02138}

\author[0000-0001-7092-2703]{Lindsay Glesener}
\affil{School of Physics and Astronomy, University of Minnesota}

\begin{abstract}
The relative amount of high temperature plasma has been found to be a useful diagnostic to determine the frequency of coronal heating on sub-resolution structures.  When the loops are infrequently heated, a broad emission measure (EM) over a wider range of temperatures is expected. A narrower EM is expected for high frequency heating where the loops are closer to equilibrium. The soft X-ray  spectrum contains many spectral lines that provide high temperature diagnostics, including lines from Fe XVII-XIX.  This region of the solar spectrum will be observed by the Marshall Grazing Incidence Spectrometer (MaGIXS) in 2020.  In this paper, we derive the expected spectral lines intensity in MaGIXS to varying amounts of high temperature plasma to demonstrate that a simple line ratio of these provides a powerful diagnostic to determine the heating frequency.  Similarly, we examine ratios of AIA channel intensities, filter ratios from a XRT, and energy bands from the FOXSI sounding rocket to determine their sensitivity to this parameter.  We find that both FOXSI and MaGIXS provide good diagnostic capability for high-temperature plasma.  We then compare the predicted line ratios to the output of a numerical model and confirm the MaGIXS ratios provide an excellent diagnostic for heating frequency.
\end{abstract}

\section{Introduction}

Since the discovery of million-degree coronal temperatures by \citet{edlen1942} and \citet{grotrian1939}, a major problem in solar physics has been to determine the mechanisms that transfer and dissipate energy into the corona.  One popular theory suggested by \cite{parker1983b,parker1983a} is that photospheric motions braid and stress magnetic field lines. The stored energy is then released through magnetic reconnection. The energy released in each reconnection event, termed as ``nanoflare'', is thought to be short-lived and occur {\em sporadically} along a single field line, because after an energy release, it requires finite time to re-build the stress of the system \citep[e.g,][]{lopez2010}.  Recent studies have shown the dissipation of Alfv\'en waves could also provide the energy required for coronal heating \citep{vanballegooijen2011,vanballegooijen2014}.  This process is also expected to produce short-lived heating events, however they would occur {\em frequently} along individual field lines \citep{asgaritarghi2012}.  Hence, the {\em frequency} of heating events on a single strand in the corona may differentiate between nanoflare and wave heating.  Unfortunately, it is well beyond the capability of any current instrument to spatially resolve single strands (widths $<$ 50\,km), though there is evidence for coherence on larger spatial scales \citep[e.g.,][]{brooks2012,brooks2013}.  The heating frequency in the highest temperature loops in the solar corona, those in the active region core, remains the most controversial \citep{tripathi2010a,tripathi2011,winebarger2011,warren2012,viall2011,viall2012,viall2013, Zanna2015}.    Even in the event of significantly improved spatial resolution to image these high temperature loops, line-of-sight confusion will still make their analysis and understanding difficult.

If the frequency of energy release on a given strand is high, the plasma along that strand does not have time to cool before being re-heated.  (Here we use the term ``strand'' to refer to the fundamental flux tube in the corona and the term ``loop'' to refer to a coherent structure in an observation.  A loop can consist of a single strand, or, more likely,  many, sub-resolution strands.)  As a result, the temperature and density of the strand remain relatively constant.  If the frequency of heating events is low (i.e., the time between two heating events on a given strand is longer than the plasma's cooling time), the plasma's density and temperature along that strand would be dynamic and evolving.  During its evolution, the temperature would be both much higher and much lower than the average temperature.  Because an observed loop is almost certainly formed of many strands \citep[see, for instance,][]{kobelski2014a}, the loop's properties may or may not reflect this plasma evolution.   In fact, if the loop is formed of many sub-resolution strands, each strand being heated randomly and then evolving, the observed loop's intensity can appear steady regardless of the dynamic nature of the plasma along a single strand \citep{klimchuk2009}.

One observation that can discriminate between low- and high-frequency heating in active region cores is the relative amount of high-temperature (${\sim}$ 5-10 MK) to average temperature plasma, which peaks at ${\sim}$ 3-5 MK.  Recent analyses have hinted at the possibility of a hot plasma component to active regions \citep{schmelz2009a,reale2009,schmelz2009b,shestov2010,testa2012,teriaca2012,brosius2014}.  Unfortunately, the ability of current instrumentation, such as {\em Hinode}'s X-ray Telescope (XRT) or Extreme-ultraviolet Imaging Spectrometer (EIS), to detect low emission measure, high-temperature plasma is limited  \citep[see e.g.,][]{winebarger2012,Testa11}.  Specifically, \cite{winebarger2012} determined that there exists a ``blind spot'' in temperature-emission measure space for {\it Hinode} XRT and EIS; they cannot detect plasma with temperatures higher than 6\,MK and emission measures lower than $10^{27}$\,cm$^{-5}$. The study of DEM in the core of active region using {\aia}, {\eis} data is also limited to cool and warm-temperature plasma \citep[see e.g.,][]{zanna2013}. Additionally, as we describe below, it is the {\it relative amount} of the plasma at these temperatures that identifies the heating frequency, not simply the presence of high temperature plasma.

One way of parameterizing the relative amount of emission at high and low temperatures is to consider the emission measure (EM) curve a broken power-law, where the emission measure increases up to the maximum temperature, i.e. $EM (T < T_m) {\sim}T^{\alpha}$, and then the emission measure decreases at larger temperature, i.e., $EM(T > T_m) {\sim}T^{-{\beta}}$.  For instance, \cite{warren2012} completed a systematic survey of 15 active region structures and determined that the EM was generally peaked at 4 MK and had a range of ${\alpha}$ from 2-5 and ${\beta}$ from 6-10.  Due to the limited high temperature constraints, though, the error on ${\beta}$ was significant, roughly 30 - 50\%.

The relationship between the emission measure and the heating frequency has been investigated in a recent series of papers \citep{bradshaw2012,reep2013a,cargill2014,barnes2016a,barnes2016b}.  \cite{bradshaw2012} and \cite{reep2013a} looked at the relationship between ${\alpha}$ and the heating frequency for low-frequency heating and nanoflare trains, respectively.  They found that they could reproduce the observed range of ${\alpha}$ in these experiments by varying heating frequency and loop length  Furthermore, \cite{cargill2014} argued that the observed range of ${\alpha}$ required the magnitude of each energy release to be proportional to the wait time between energy events.   These papers suggest that measuring the relative amount of ${\sim}$ 1  MK emission to ${\sim}$ 4 MK emission (i.e., ${\alpha}$) could be used to constrain the heating frequency and the magnitude of the energy release.   However, there is generally more cool plasma in the coronal arcade surrounding the hot core and the footpoints of high temperature loops form a bright reticulated pattern called moss, meaning well isolating the cool emission from a single structure can be difficult.

\cite{barnes2016a} and \cite{barnes2016b} investigated the high temperature emission expected with different heating frequencies for low-frequency nanoflares and nanoflare trains, respectively.  They find many more factors impact the relative amount of high temperature emission, including ionization non-equilibrium and differential heating between the ions and electrons.  They argue that the high temperature emission can not be fit by a simple power law as generally the fall off increases as the temperature increases, meaning ${\beta}$ depends upon the temperature range over which it is measured (see Figure~7 in \citealt{barnes2016b}).  They argue instead that ratios of spectral lines formed at different temperatures may be a better diagnostic.  The ratio of Fe XII to Fe XIX spectral lines observed by the Extreme Ultraviolet Normal Incidence Spectrograph (EUNIS-13) was used to argue that EUNIS-13 observation supported nanoflare heating \citep{brosius2014}.  Though \cite{barnes2016b} calculates the expected ratio as a function of heating frequency for a variety of line pairs, they always include a lower temperature line, such as Fe XII or Fe XV, which will be formed at or below the peak of the emission measure curve, and hence can be impacted by both ${\alpha}$ and ${\beta}$.

In this paper, we investigate whether line ratios (or channel/filter/energy ratios) are sensitive to the fall off of the high temperature emission while being insensitive to the lower temperature emission.  We use a simple broken power law as an approximation for the emission measure curve and choose a range of ${\alpha}$ and ${\beta}$ consistent with \cite{warren2012}. The fact is that the observation of spectral lines strongly depend on the value of peak emission measure, which in turn dictates the statistical uncertainty. We establish the strong dependence of various line intensity ratios on ${\beta}$ by considering statistical and instrumental uncertainties. We first consider the expected observations from the Marshall Grazing Incidence X-ray Spectrometer (\magixs), which will observe the soft X-ray spectrum from 6-24\,\AA.  We calculate the expected line ratios from the strongest Fe XVII, Fe XVII, and Fe XIX lines in the \magixs\ wavelength range and find they are sensitive to ${\beta}$ and insensitive to ${\alpha}$.  We perform similar analysis with two energy passbands of the {\foxsi} sounding rocket instrument and find it is similarly sensitive to the high temperature fall off and insensitive to the low-temperature fall off.  Finally, we consider {\aia} and {\xrt} and find they are both insensitive to ${\beta}$ or sensitive to both ${\alpha}$ and ${\beta}$ making such ratios difficult to interpret.  We argue that instruments sensitive to the high temperature emission alone, like \magixs\ or \foxsi, offer unique observations to constrain the frequency of heating in the solar corona.

\section{Method}

\begin{figure}[h]
    \centering
    \includegraphics[width=0.7\linewidth]{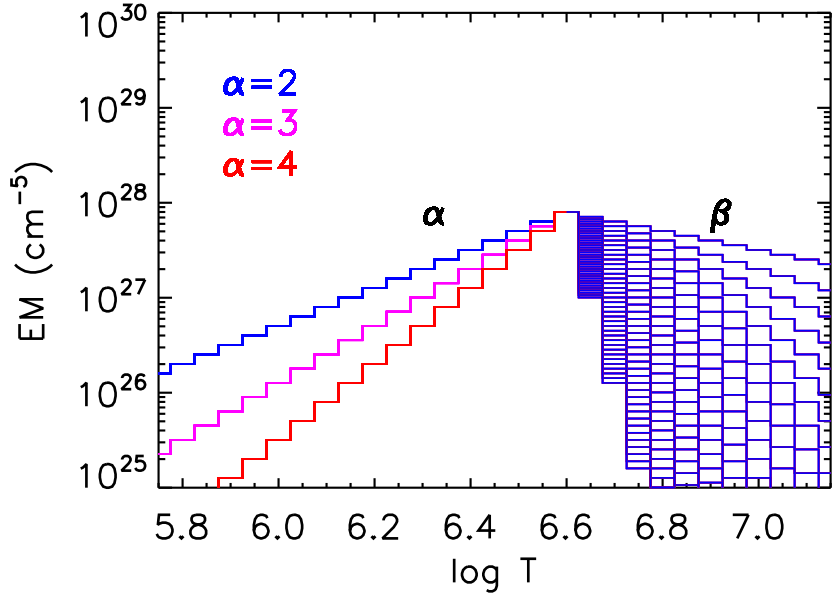}
    \caption{The synthetically constructed EM distributions characterized by two slope parameters. The positive slope ${\alpha}$ defines the low temperature EM distribution between Log T = 6.0 and 6.6; the negative slope  ${\beta}$ defines the high temperature EM distribution between Log T = 6.6 and 7.1. Range of ${\alpha}$'s and  ${\beta}$'s are obtained from \citep{warren2012}.}
    \label{fig:EMslopes}
\end{figure}

In this section, we present our method to test the sensitivity of various instruments to the high temperature EM slope.  First, we establish a variety of emission measure curves. We assume the EM curve can be well modeled as a broken power law, i.e.,
\begin{eqnarray}
EM(T < T_m) &=& EM_0 T^{\alpha}\nonumber \\
EM (T> T_m) &= & EM_0T^{-{\beta}}.
\label{equation2}
\end{eqnarray}
\cite{warren2012} found that all the active region structures had emission measure curves that peaked at Log $T_m$ = 6.6 ($T_m {\sim}$ 4\,MK); for this paper, we use this value as the peak of the emission measure curves.  We have the same peak emission measure ($8 {\times}10^{27}$ cm$^{-5}$) for all emission measure curves in this paper, this is consistent with the Active Region \#8 from \cite{warren2012}, which is a medium sized AR corresponding to median AR in the survey. We choose a distribution of ${\alpha}$ and ${\beta}$ limited by the distribution measured by \cite{warren2012}; we use three ${\alpha}$'s (i.e., ${\alpha}$  = 2, 3 and 4) and a range of ${\beta}$ values from 1 through 20.   All emission measure curves considered in this analysis are shown in Figure~\ref{fig:EMslopes}, which is produced using  equation~\ref{equation2} with a fixed peak EM value at Log $T_m$ = 6.6 ($T_m {\sim}$ 4\,MK).

 Using these emission measure curves, we then calculate the expected intensity in the selected spectral lines or channels, and find the relationship between the ratio of two intensities to ${\alpha}$ and ${\beta}$. To calculate the intensities, we use the emissivity function of the spectral line or the temperature response of the instrument for a given channel and fold through the synthetic EM distributions following
 \begin{equation}
    I_{{\lambda}~or~E}^{{\alpha},{\beta}} = {\sum}~EM(T)^{{\alpha},{\beta}} R(T)_{{\lambda}~or~E}
    \label{eq:context_eq}
\end{equation}
where $R(T)_{{\lambda}~or~E}$ is the emissivity function of a spectral line or the temperature response function of an instrument.  To understand the limitations of each instrument on its sensitivity to ${\alpha}$ and ${\beta}$, we use realistic exposure times and include the photon noise in the calculation.  The detailed calculations for each instrument are given below.

\section{Results}
\label{sec:result}
\subsection{MaGIXS}

The Marshall Grazing Incidence X-ray Spectrometer (\magixs) is an instrument being developed by the Marshall Space Flight Center (MSFC) and the Smithsonian Center for Astrophysics (SAO) to be flown on a sounding rocket in April 2020.  \magixs\ will observe the soft X-ray spectrum from 24 - 6.0 \AA (0.5 - 2.0 keV), with a spatially resolved component along an 8 arcmin long, 2.5 arcsecond wide slit. The nominal spectral resolution of MaGIXS is 0.022\,\AA\ with a plate scale of 0.011\,\AA/pixel in the dispersion direction and the spatial resolution along the slit is 6\arcsec ~with 2.8\,\arcsec/pixel.  It is a grazing incidence imaging spectrometer that can be described by two primary subsystems: a Wolter-I telescope, with a slit jaw (context) imager, and a soft X-ray spectrometer. The slit is placed at the focal plane of a single shell, nickel-replicated Wolter-I telescope. On the backside of the slit is a finite conjugate mirror pair, a blazed, varied-line space reflective grating, and a CCD detector. The finite conjugate mirror pair are identical nickel-replicated paraboloidal mirror shells (SM1 and SM2, respectively), which re-image the slit from the rear. Between SM2 and the detector sits the grating that will diffract a converging (focused) cone of rays to the detector, positioned off-axis. The detector will capture the first order diffraction of the image of the slit \citep{doi:10.1117/12.856793,doi:10.1117/12.2313997,doi:10.1117/12.2232820}.

The Wolter-I telescope and paraboloidal mirror pair were designed and fabricated by MSFC, using the same electroform nickel-replication techniques as developed for instruments such as Astronomical Roentgen Telescope X-ray Concentrator ({\it ART-XC}; \citealt{doi:10.1117/12.2027141, doi:10.1117/12.2056595, 2017ExA....44..147K}), the Focusing Optics X-ray Solar Imager ({\it FOXSI};  \citealt{doi:10.1117/12.827950,doi:10.1117/12.895271,doi:10.1117/12.2024277}), and Imaging X-ray Polarimetry Explorer ({\it IXPE}; \citealt{stephen2018}). 
For these instruments, a electroless nickel-coated mandrel is fabricated and polished to a required optical prescription. The mandrel is then replicated, creating one or more thin shells.  However, since the desired angular resolution for {\magixs} is significantly smaller than the angular resolution achieved for these other projects, a state of the art computer-numerical-control (CNC) polishing technique was employed on the {\magixs} mandrels. The figures of both the Wolter-I mandrel and the single paraboloid spectrometer mandrel were deterministically polished to a fine figure, with slope errors down to just a few arcseconds, at spatial wavelengths $>$5\,mm. Great progress was made at reducing these low-frequency, axial figure errors over a designated 100 degree region of both mandrels. A complete analysis of the MaGIXS replicated mirror performances is compiled in \citep{champey2019}.

\begin{deluxetable}{c c c}
\tablecaption{Strong Lines in the MaGIXS Wavelength Range \label{tab:lines}}
\tablehead{
\colhead{Ion} & \colhead{Wavelength}  & \colhead{Log Maximum Temperature}    }
\startdata
Fe \sc{XVII}& 15.01\,\AA & 6.6 \\
Fe \sc{XVIII}& 14.21\,\AA & 6.8 \\
Fe  \sc{XIX} & 13.53\,\AA & 6.95 \\
\enddata
\end{deluxetable}

\begin{figure}[h]
    \includegraphics[width=0.5\linewidth]{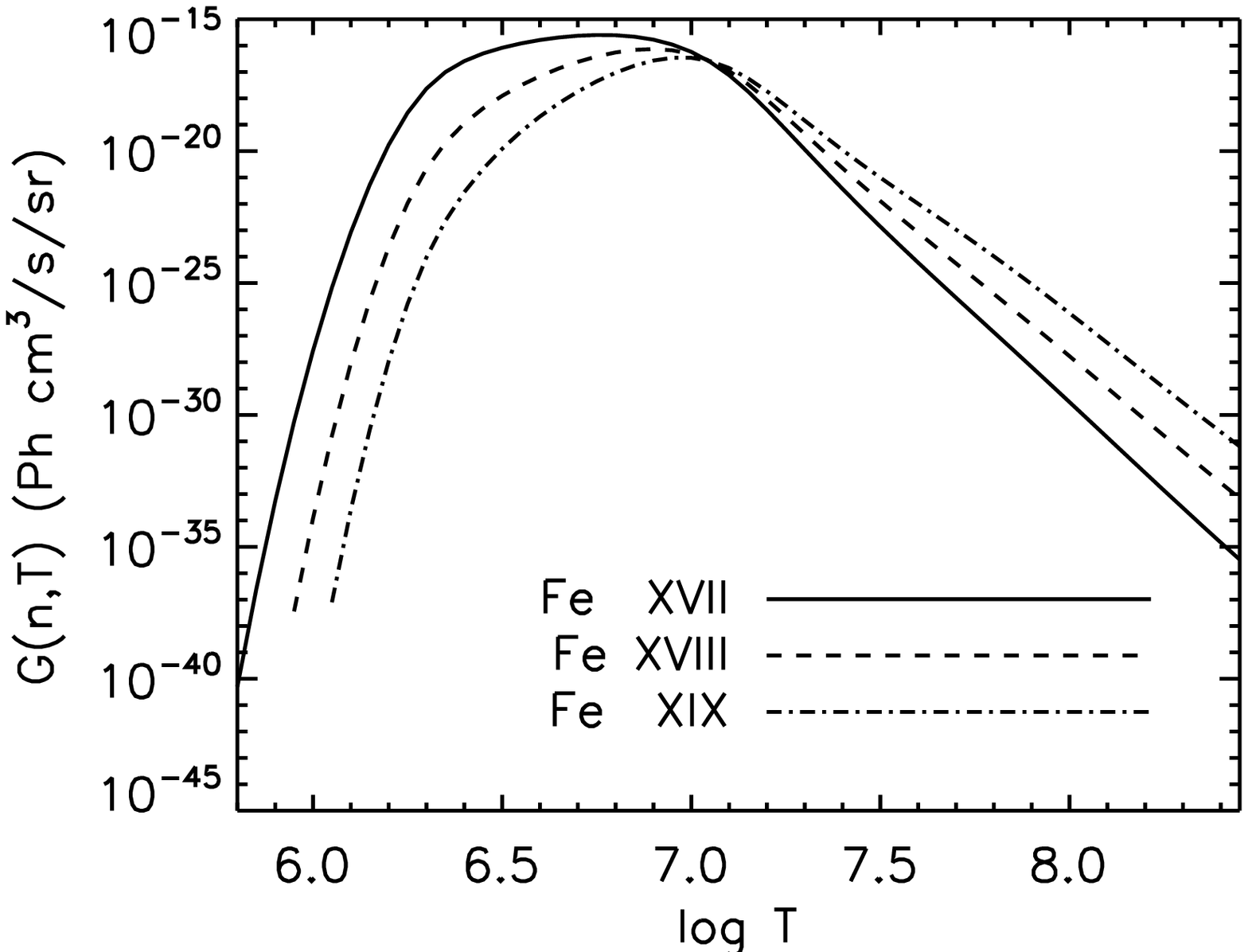}
    \includegraphics[width=0.5\linewidth]{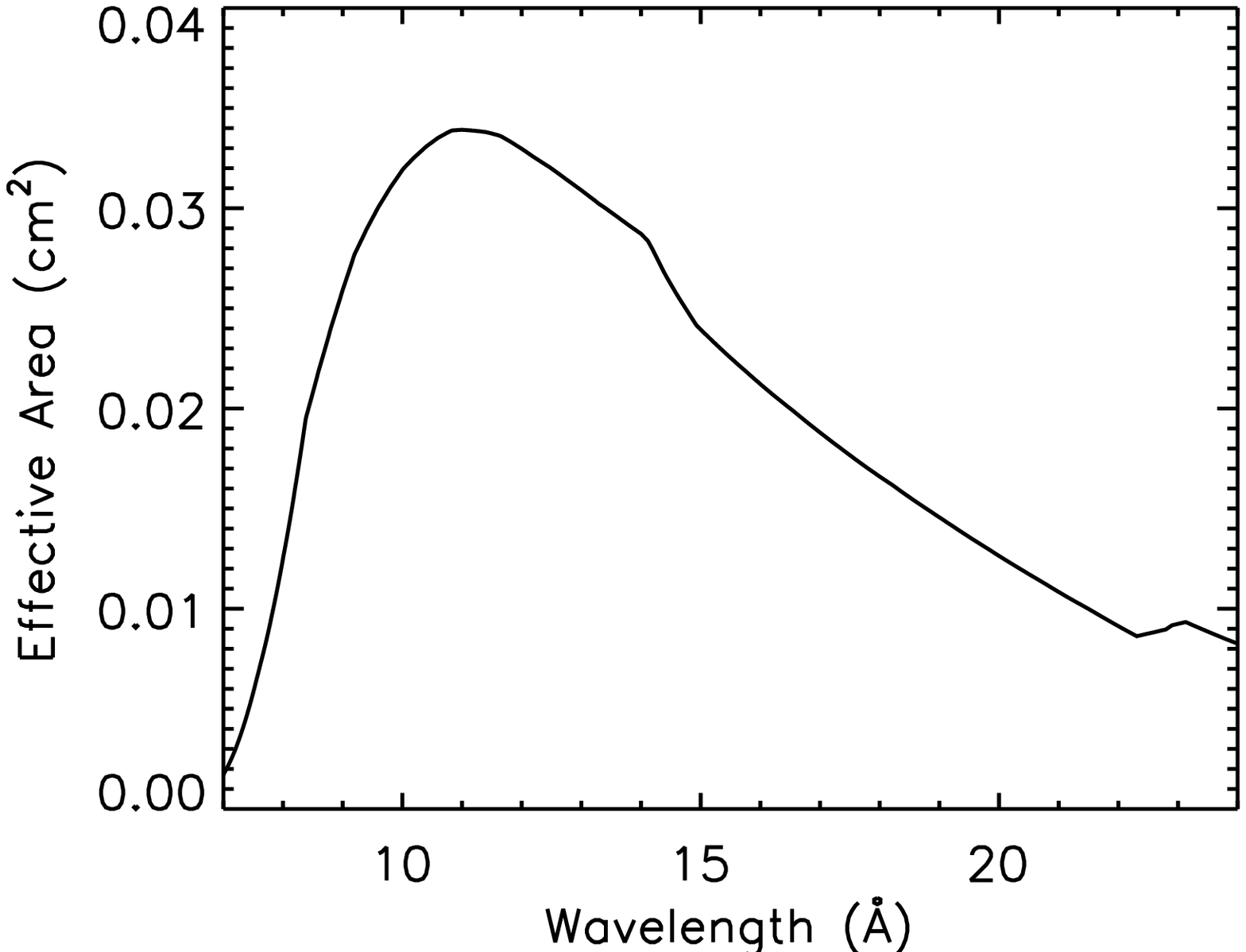}
    \caption{(Left) The expected line intensity per unit emission measure (or) the emissivity function for Fe XVII, Fe XVIII  and  Fe XIX, calculated using CHIANTI atomic database version 8.0.7 with coronal abundances \citep{schmelz2012} and ionization equilibrium \citep{delzanna_etal:2015}. (Right) The effective area of the {\magixs} instrument derived analytically.}
    \label{fig:gofnt}
\end{figure}

Table~\ref{tab:lines} gives some of  the strongest Fe-ion lines in the {\magixs} wavelength range.  Because these spectral lines are formed at or above the peak of the emission measure curve, their intensity is expected to be strongly dependent on  ${\beta}$ and less sensitive or insensitive to ${\alpha}$.   To calculate the intensity in these spectral lines for the variety of EM curves, we first obtain the contribution function ($G(T)$) using CHIANTI atomic database version 8.0.7; these are shown in the left panel of Figure~\ref{fig:gofnt}.  Here, we assume coronal abundances \citep{schmelz2012} and the standard CHIANTI ionization equilibrium \citep{delzanna_etal:2015}.  We then calculate the expected synthetic line intensity using the Equation~\ref{eq:context_eq}, where the resulting intensity is the expected intensity in units of photons/s/cm$^2$/sr. To generate the predicted counts to be observed during {\magixs} flight in ``real'' units, we multiply $I_{\lambda}$ with the {\magixs} effective area at that wavelength, the steradians/spatial pixel, and the rocket flight duration (300\,s) to yield expected photons/spectral line/spatial pixel along the slit. The effective area curve of {\magixs} derived analytically is shown in the right panel of Figure~\ref{fig:gofnt}.

\begin{figure}[h]
   \includegraphics[width=0.34\linewidth]{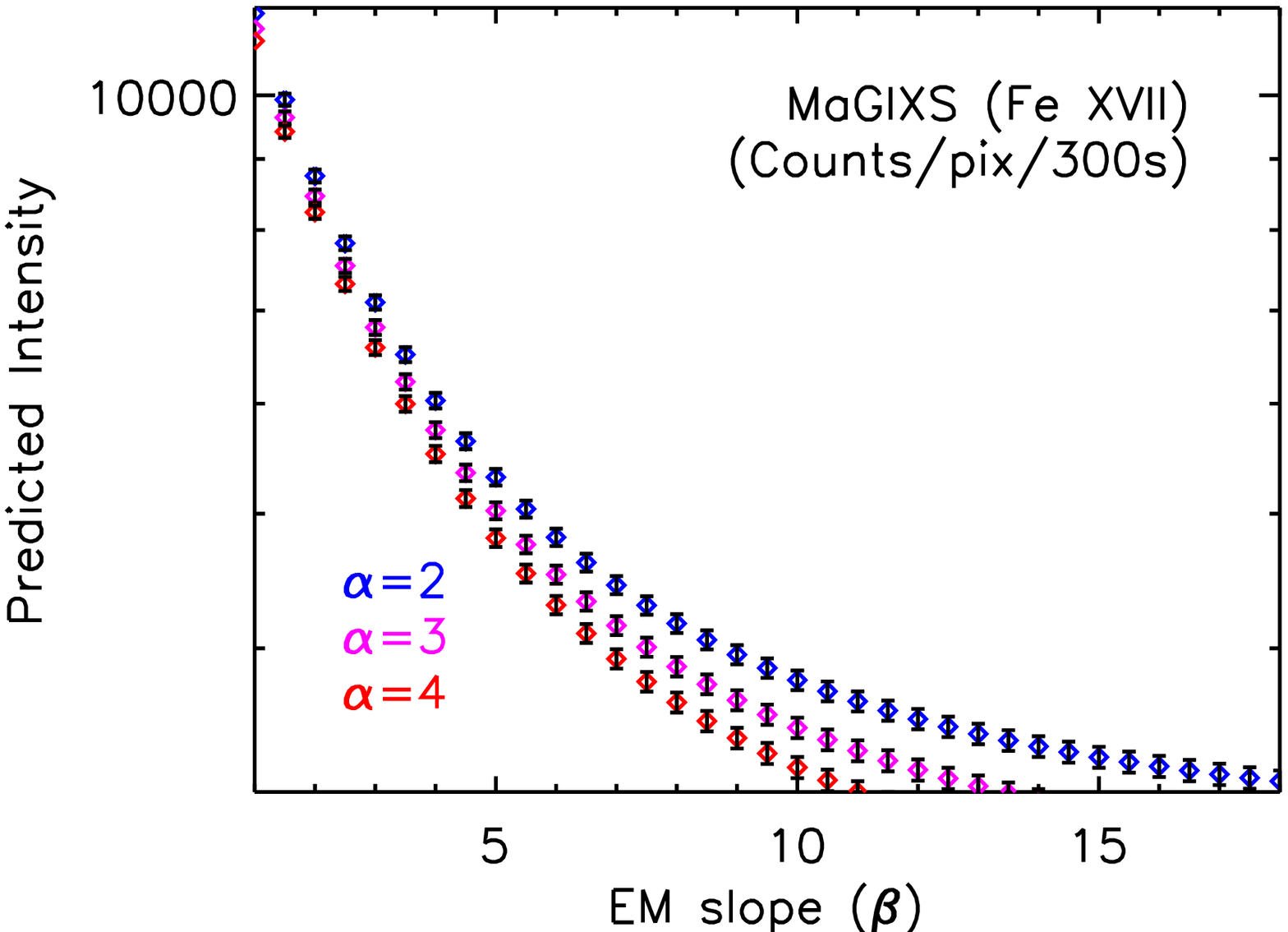}
    \includegraphics[width=0.34\linewidth]{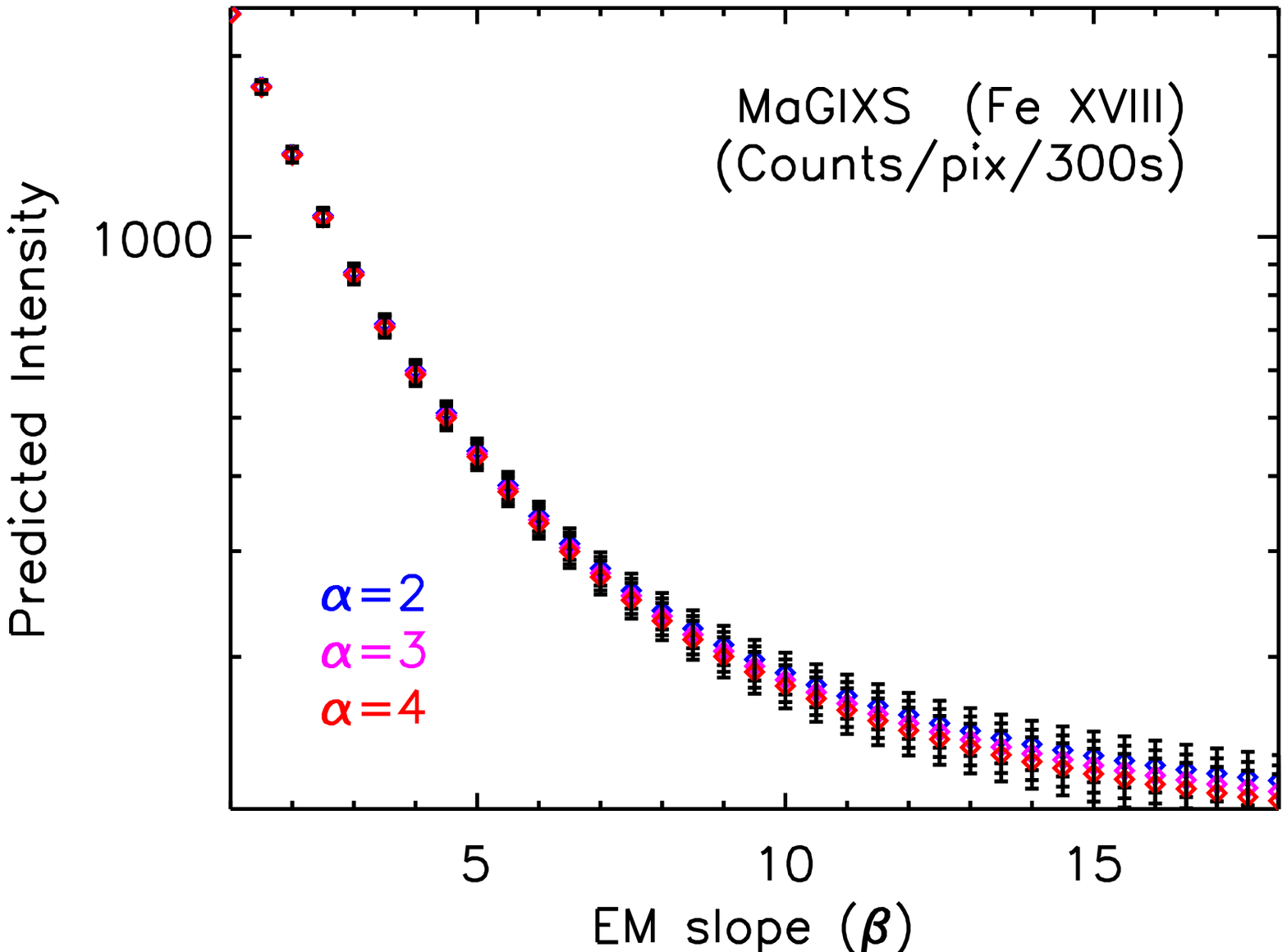}
    \includegraphics[width=0.34\linewidth]{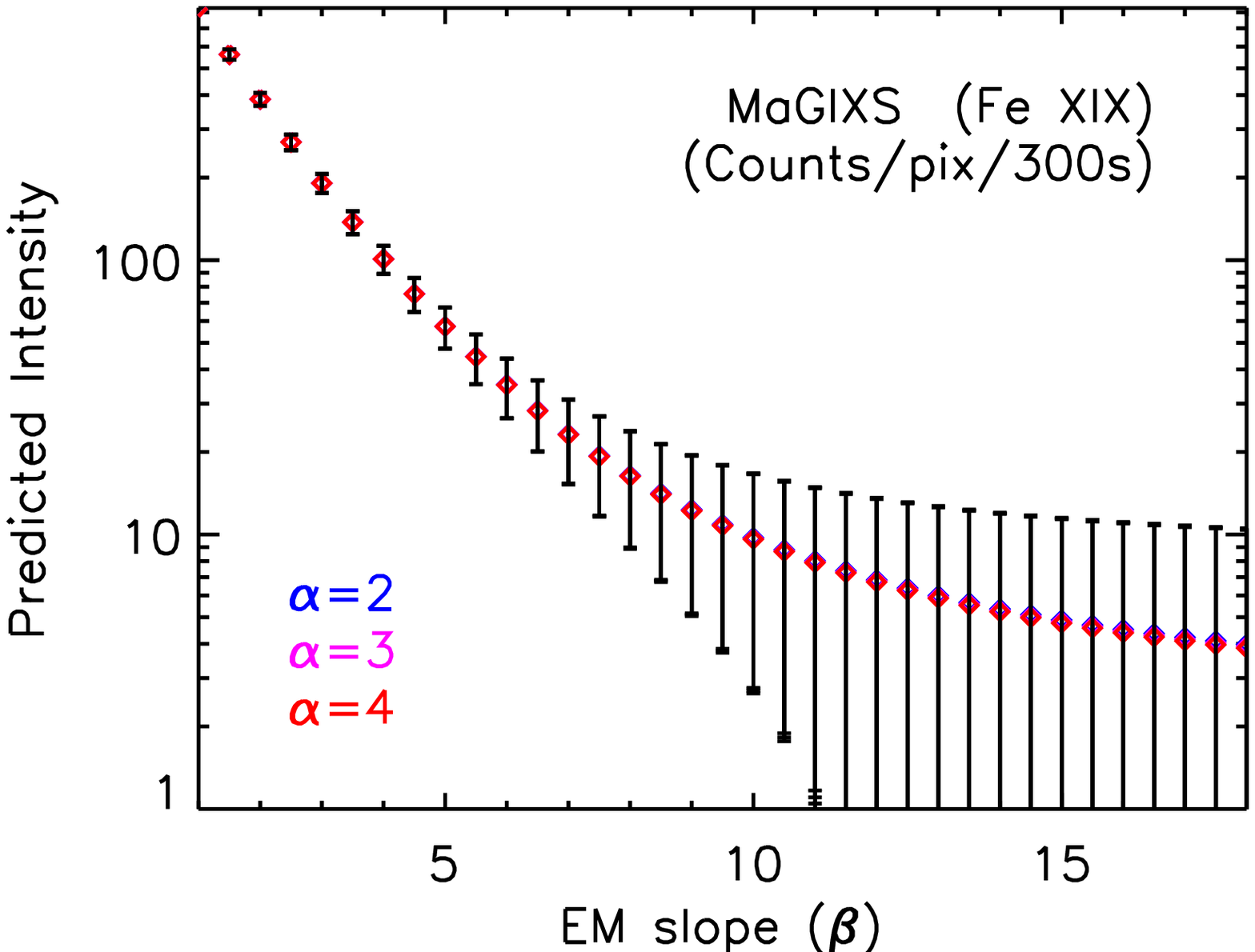}
    \caption{The expected signal to be observed in {\magixs} at selected wavelengths (Fe XVII, Fe XVIII and Fe XIX) plotted as a function of EM slope ${\beta}$ on the X-axis. Overplotted in three different colors represent the expected signal for EM distributions with different ${\alpha}$.  The predicted intensity of high temperature ions Fe XVIII and Fe XIX seems to be insensitive to  ${\alpha}$ and falls off  sharply  with increasing negative slope ${\beta}$. The predicted intensity of Fe XVII is sensitive for both ${\alpha}$ and ${\beta}$ as expected.} The error bars include statistical noise and readout noise added in quadrature.
    \label{fig:ph_predicted_magixs}
\end{figure}

 Figure~\ref{fig:ph_predicted_magixs} shows the predicted {\magixs} intensities in units of photon counts for each spectral line in a single pixel along the slit, plotted as a function of EM slope ${\beta}$. Overplotted in different colors are the intensities with different EM slope  ${\alpha}$. The error bars represent Poisson uncertainties arising from photon counting statistics quadrature summed with the detector's readout noise, required  to be $<$25 e$^{-}$ rms/pix/readout.  The {\magixs} camera uses the same CCD and readout electronics of the High Resolution Coronal Imager (Hi-C 2.1) camera, which achieved $< 10$ e$^{-}$ rms/pix/readout noise (Rachmeler et al., in preparation).

\begin{figure}[h]
    \includegraphics[width=0.34\linewidth]{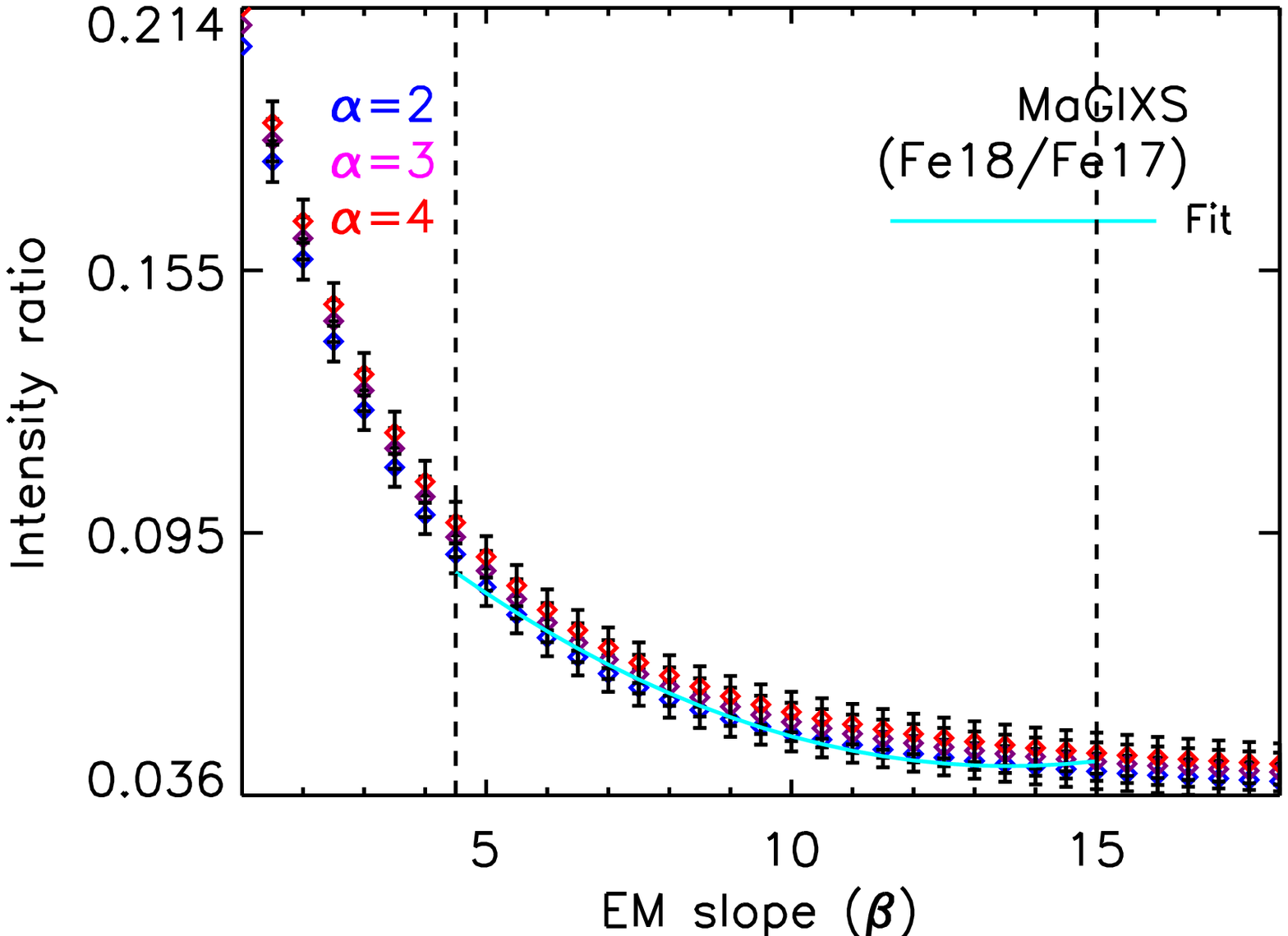}
        \includegraphics[width=0.34\linewidth]{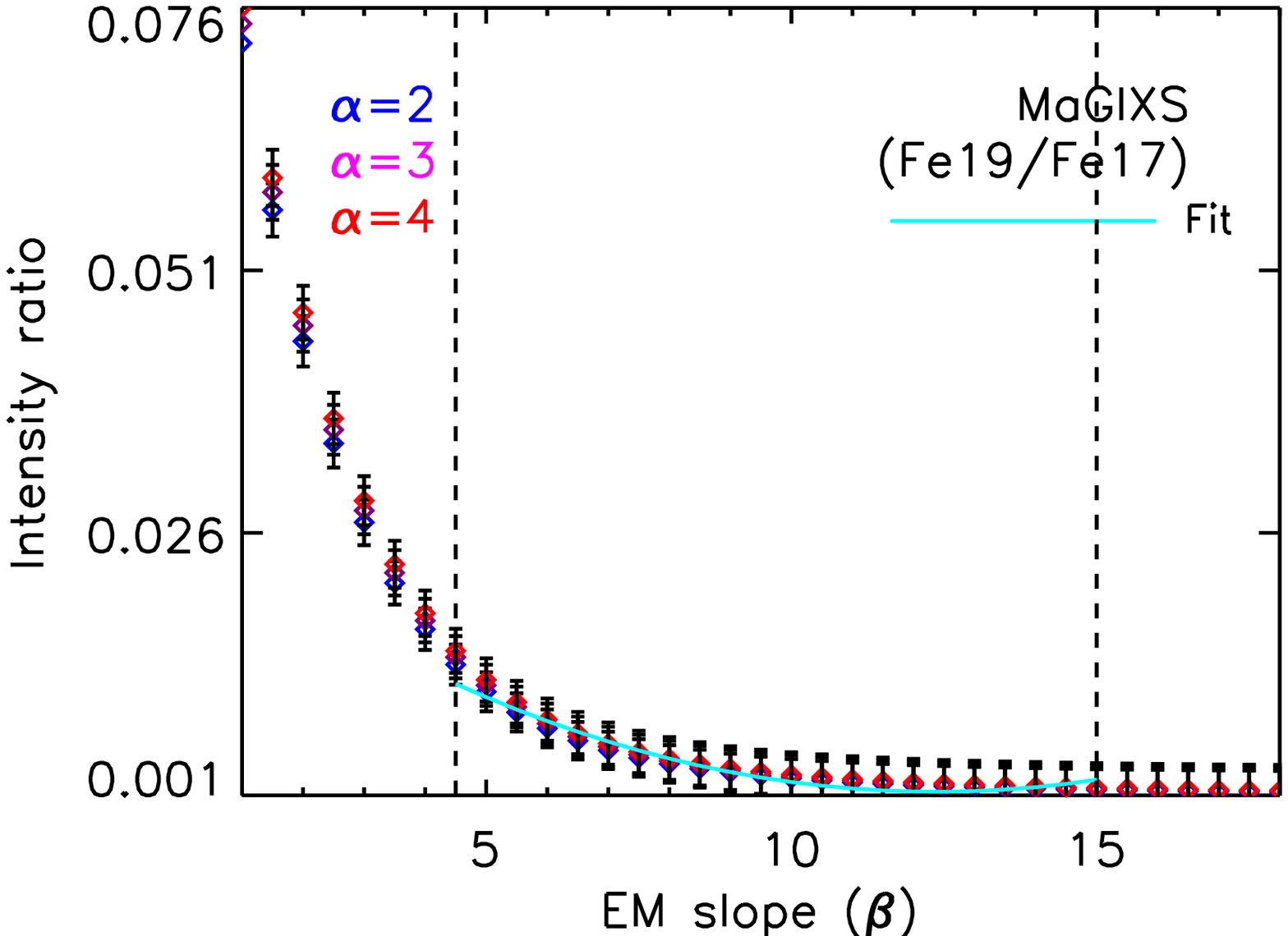}
            \includegraphics[width=0.34\linewidth]{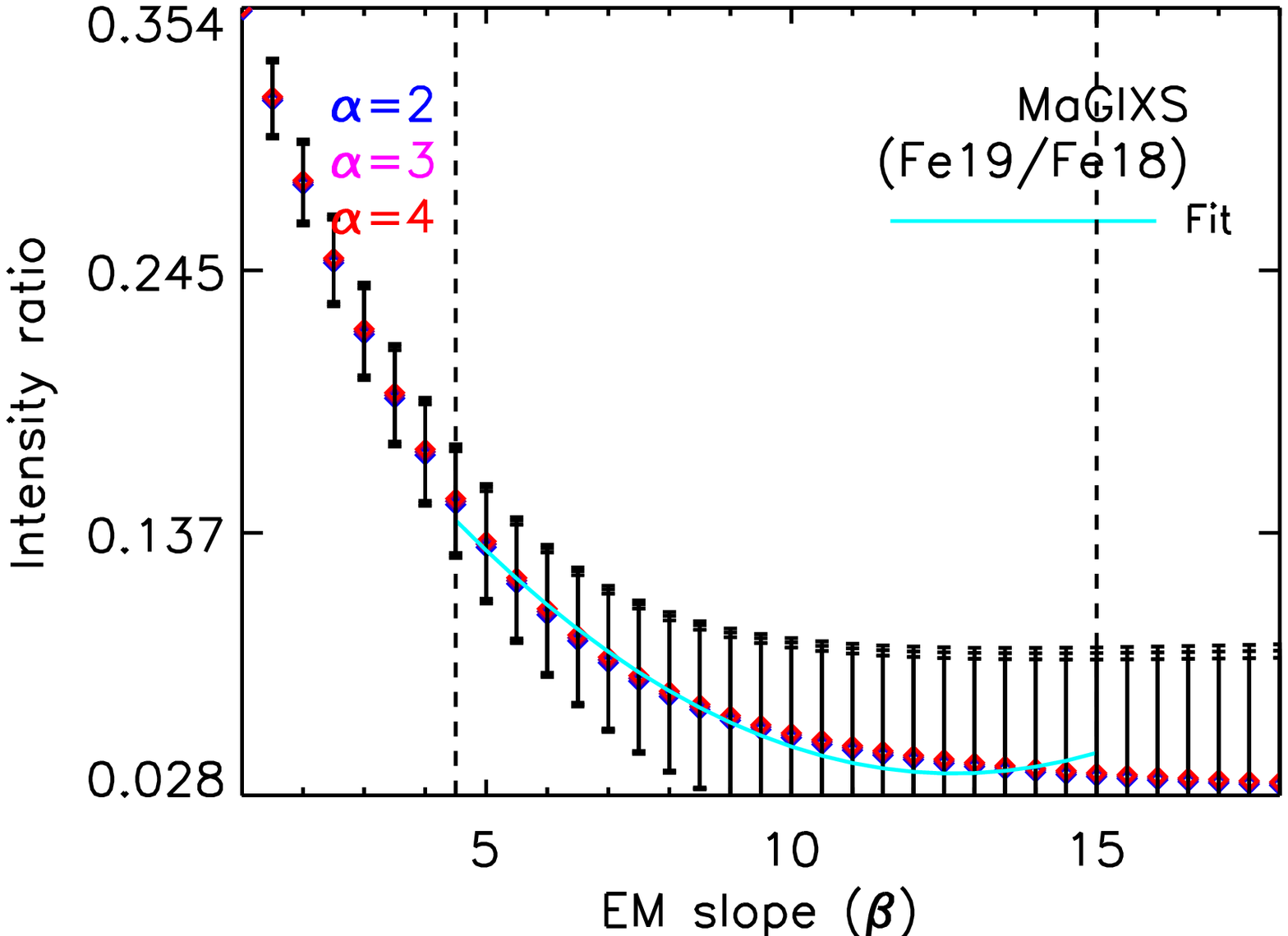}
            \caption{The ratio between two {\magixs} line intensity plotted as a function of EM slopes. The different colors denote different ${\alpha}$ values. The error bars are propagated uncertainties from predicted intensity. The ratios are fitted with a second order polynomial (solid line) within the ${\beta}$ range between two vertical dashed lines.}
    \label{fig:ratiovsslope}
  \end{figure}

 From Figure \ref{fig:ph_predicted_magixs}, we observe that the intensity of the spectral lines emitted by the hotter ions (Fe XVIII and Fe XIX) are insensitive to   ${\alpha}$, whereas Fe XVII line intensity is sensitive to  ${\alpha}$ and ${\beta}$. The expected intensity falls sharply with increasing ${\beta}$ for Fe XVIII and Fe XIX as compared to Fe XVII.   To investigate how the predicted intensities could be useful to determine the hot EM slope   ${\beta}$, we calculate the line intensity ratios and plot them as a function of EM slope as shown in Figure~\ref{fig:ratiovsslope}. The intensity ratio plots of the selected {\magixs} lines are fairly insensitive to   ${\alpha}$, and vary strongly as a function of   ${\beta}$.

\begin{figure}[h]
    \includegraphics[width=0.34\linewidth]{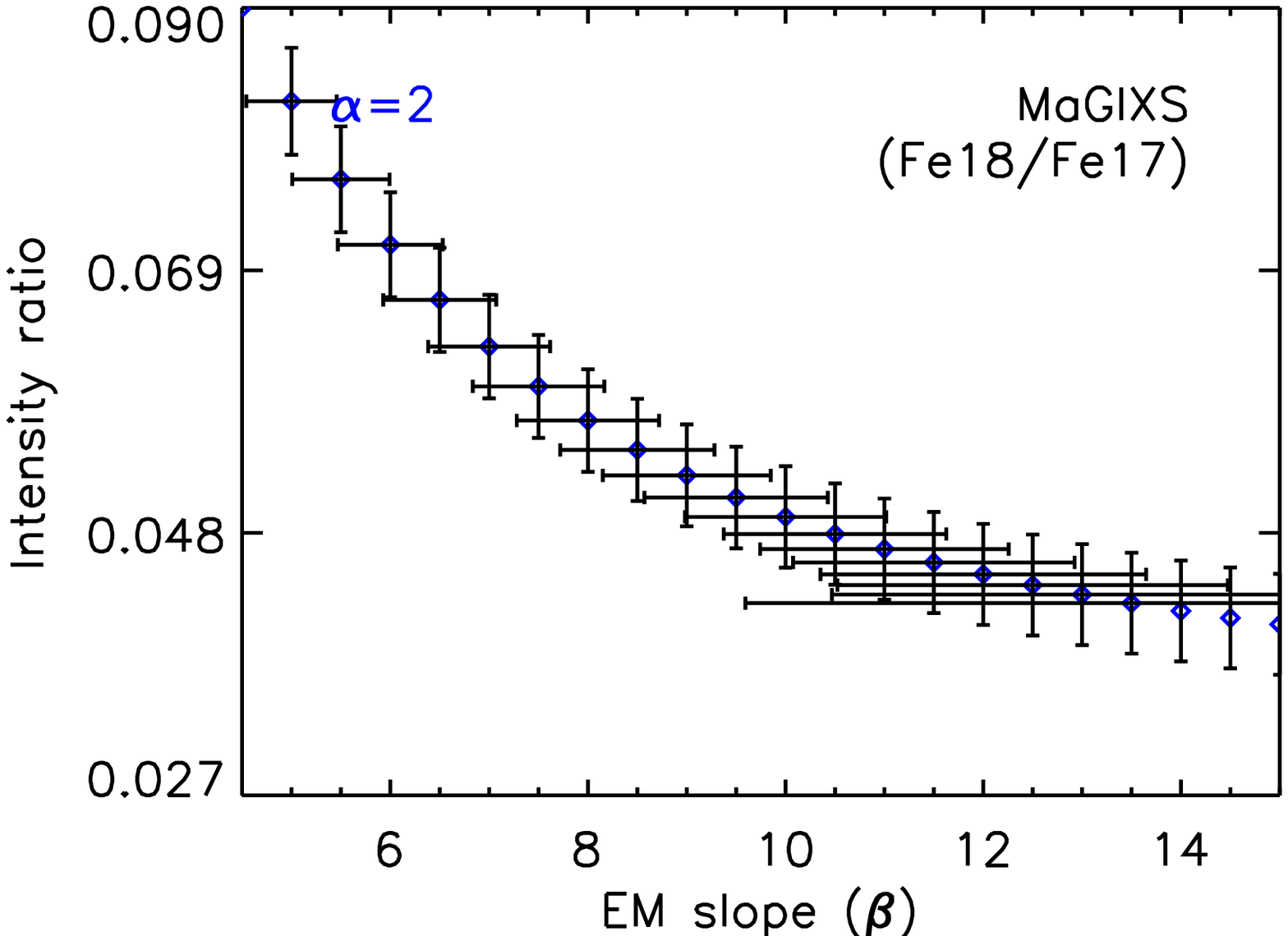}
        \includegraphics[width=0.34\linewidth]{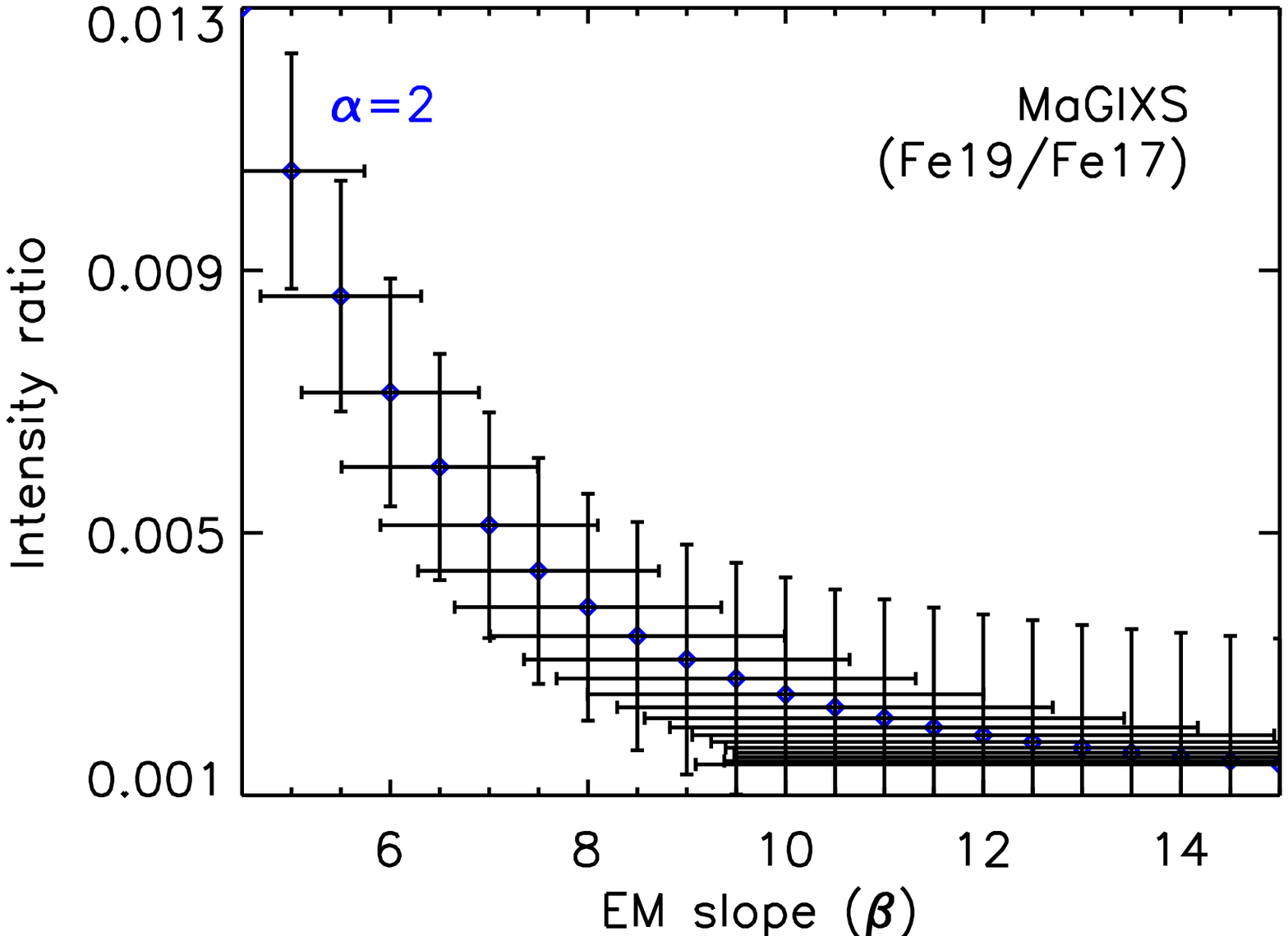}
            \includegraphics[width=0.34\linewidth]{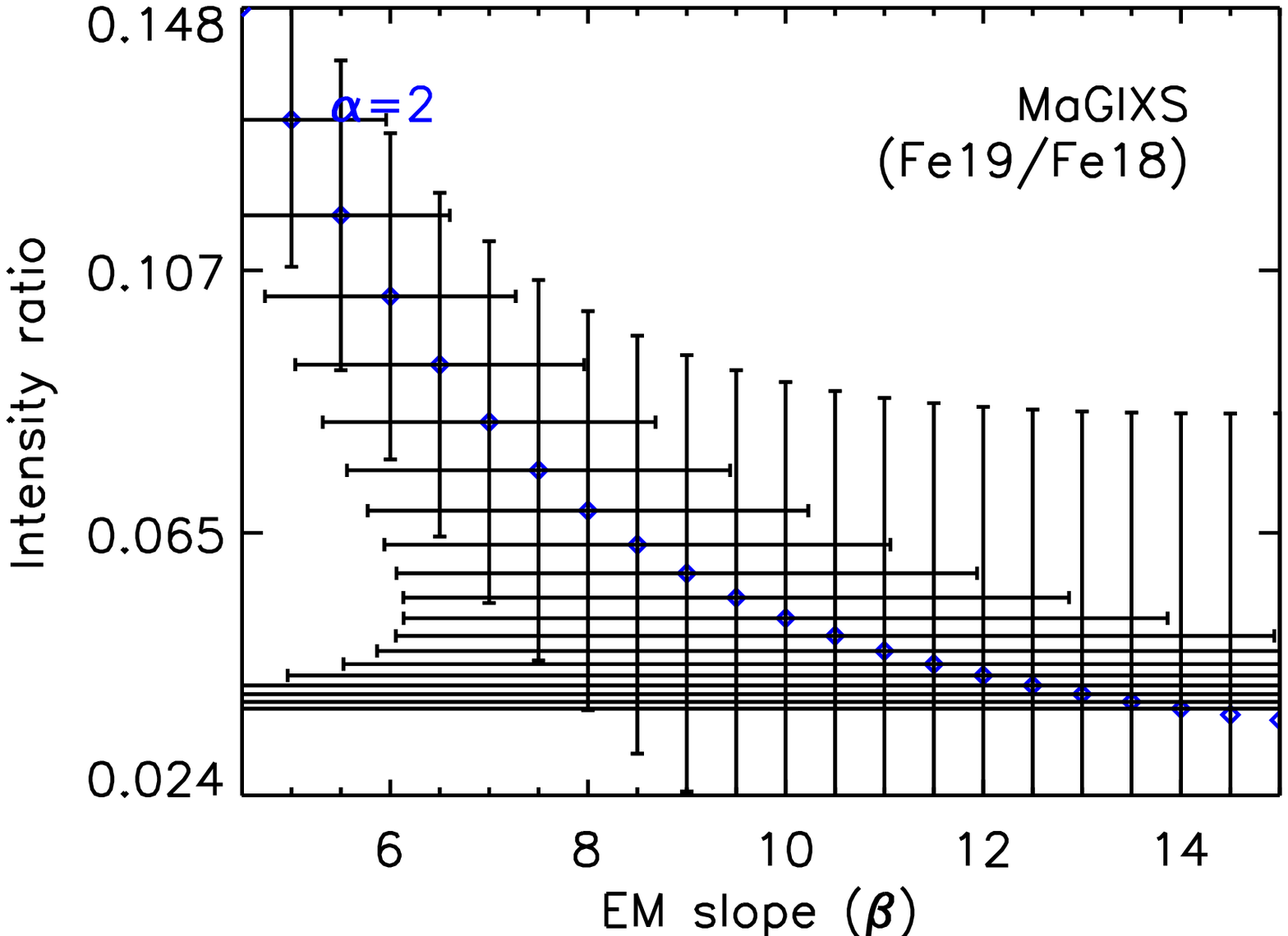}
            \caption{The ratios between {\magixs} line intensity plotted as a function of EM slopes with error bars in both ratio and ${\beta}$. The error in ${\beta}$ is obtained by inverting the empirical fit and propagating error in the respective ratios.}
    \label{fig:ratiovsslope_error}
  \end{figure}

Next, we estimate the uncertainty (${\sigma}_{\beta}$) in the EM slope from the uncertainty in the different {\magixs} line intensity ratios.  We empirically fit line intensity ratios-vs-${\beta}$ using polynomial of second order as shown  in  Figure~\ref{fig:ratiovsslope}. For fitting, we considered range of observed ${\beta}$'s from 4 to 15 \citep{warren2012}.  We then invert the quadratic equation and propagate error in the intensity ratios to derive the uncertainty in the EM slope (${\sigma}_{\beta}$).  Figure \ref{fig:ratiovsslope_error} shows the error bars in both ratio and ${\beta}$ and Table~\ref{tab:errorinbeta} gives the range of uncertainty (${\sigma}_{\beta}$) in the EM slope derived using different {\magixs} line intensity ratios. The typical values for ${\sigma}_{\beta}$/${\beta}$ to be observed from {\magixs} will be more tightly constrained than the range of values obtained before, which can be up to 50\% \citep{warren2012}.

\begin{deluxetable}{c c c}
\tablecaption{Range of uncertainty (${\sigma}_{\beta}$) in the estimated EM slope using different {\magixs} line intensity ratios \label{tab:errorinbeta}}
\tablehead{
\colhead{Line ratio} & \colhead{Error in ${\beta}$ (${\sigma}$$_{\beta}$)}  & \colhead{Error in ${\beta}$ (${\sigma}$$_{\beta}$)}\\
 & 4 $<$ ${\beta}$ $<$ 10 & 10 $<$ ${\beta}$ $<$ 16}
\startdata
Fe~{\sc XVIII}~/~Fe~{\sc XVII}& 0.5 - 1.0 & 1.0 - 3.0 \\
Fe~{\sc XIX}~/~Fe~{\sc XVII}& 0.6 - 2.0& 2.0 - 4.0 \\
Fe~{\sc XIX}~/~Fe~{\sc XVIII} & 1.0 - 4.0 &4.0 - 14.0  \\
\enddata
\end{deluxetable}

\subsection{FOXSI-2}

The Focusing Optics X-ray Solar Imager \citep[\it FOXSI,][]{krucker2009,  krucker2013, Krucker2014} is a hard X-ray experiment successfully flown on three sounding rocket campaigns, acquiring observations of the Sun using focusing optics in the 4~-~20 keV energy range with an energy resolution of ${\sim}$ 0.5 keV \citep{Ishikawa2011, Athiray2017}. The design involves seven Wolter~-~I type optic modules each paired with a semiconductor photon counting detector. The {\it FOXSI} energy bands are sensitive to observe high temperature plasma above 5 MK.  The second rocket flight campaign ({\foxsi}) \citep{Christe2016, Glesener2016}  observed quiet and microflaring ARs demonstrating the measurement of a well-constrained faint emission at high temperatures up to 10 MK (\citealt{Ishikawa2017}, Athiray et al., to be submitted).

\begin{figure}[h]
    \includegraphics[width=0.34\linewidth]{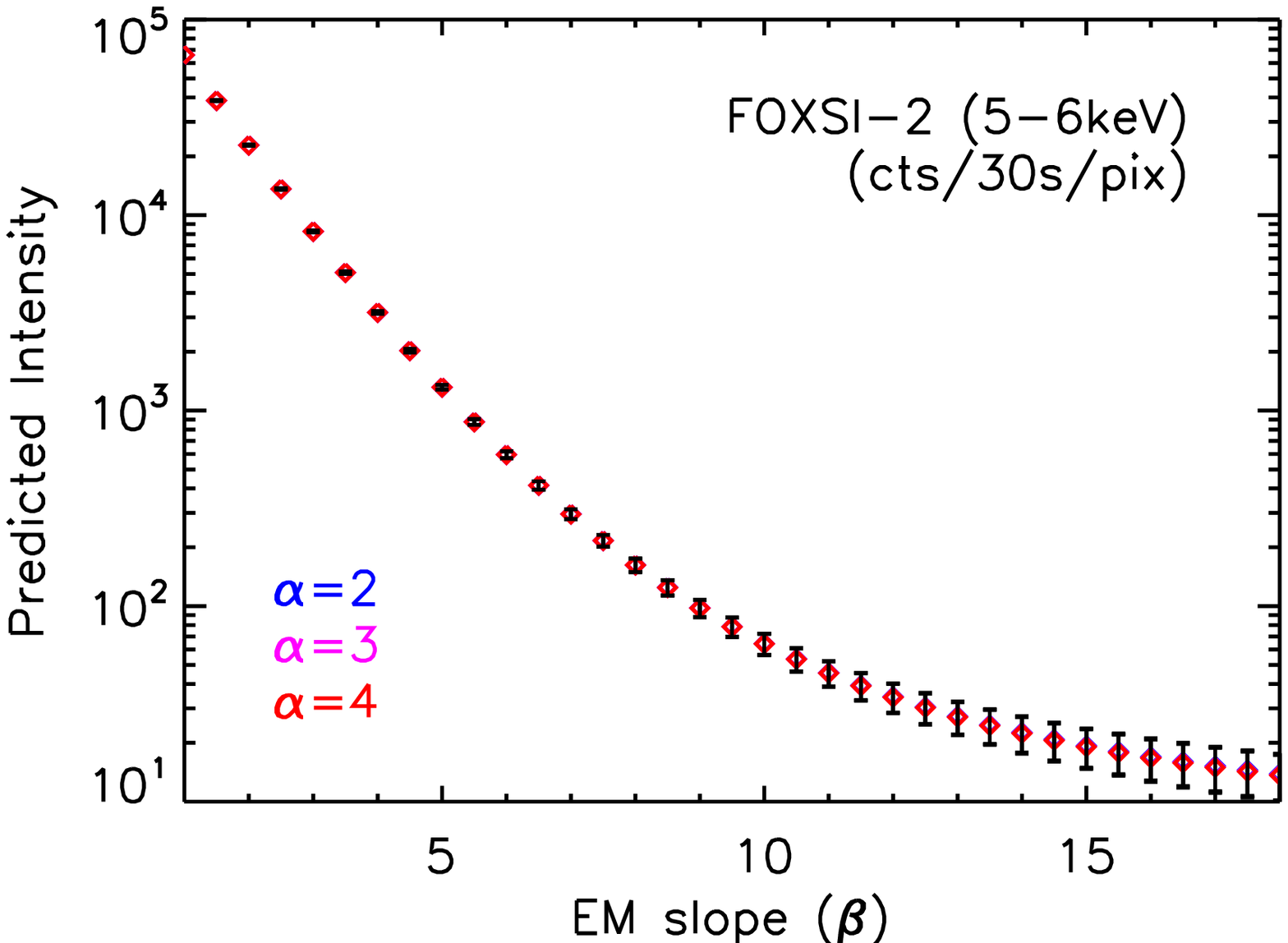}
    \includegraphics[width=0.34\linewidth]{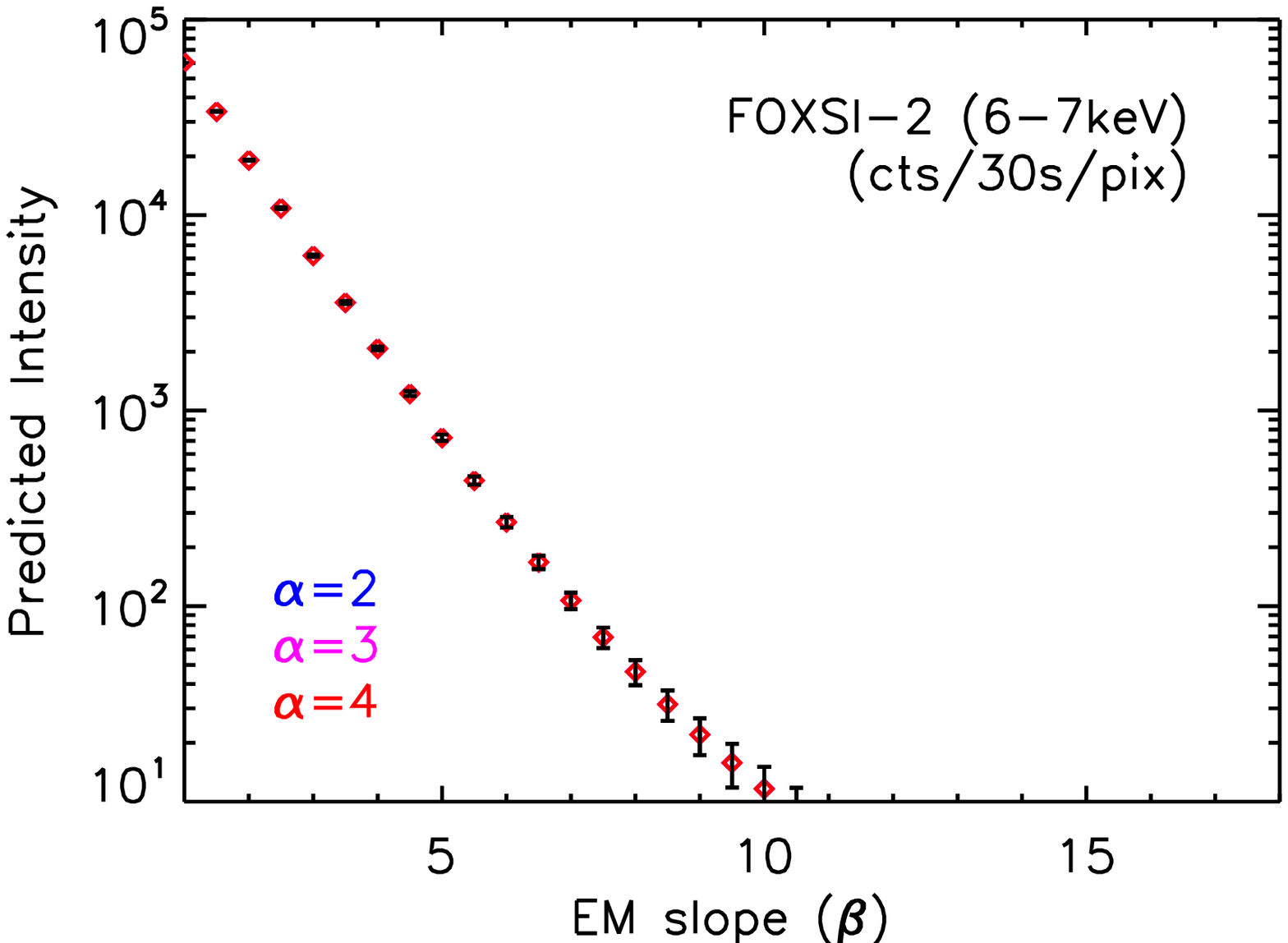}
    \includegraphics[width=0.35\linewidth]{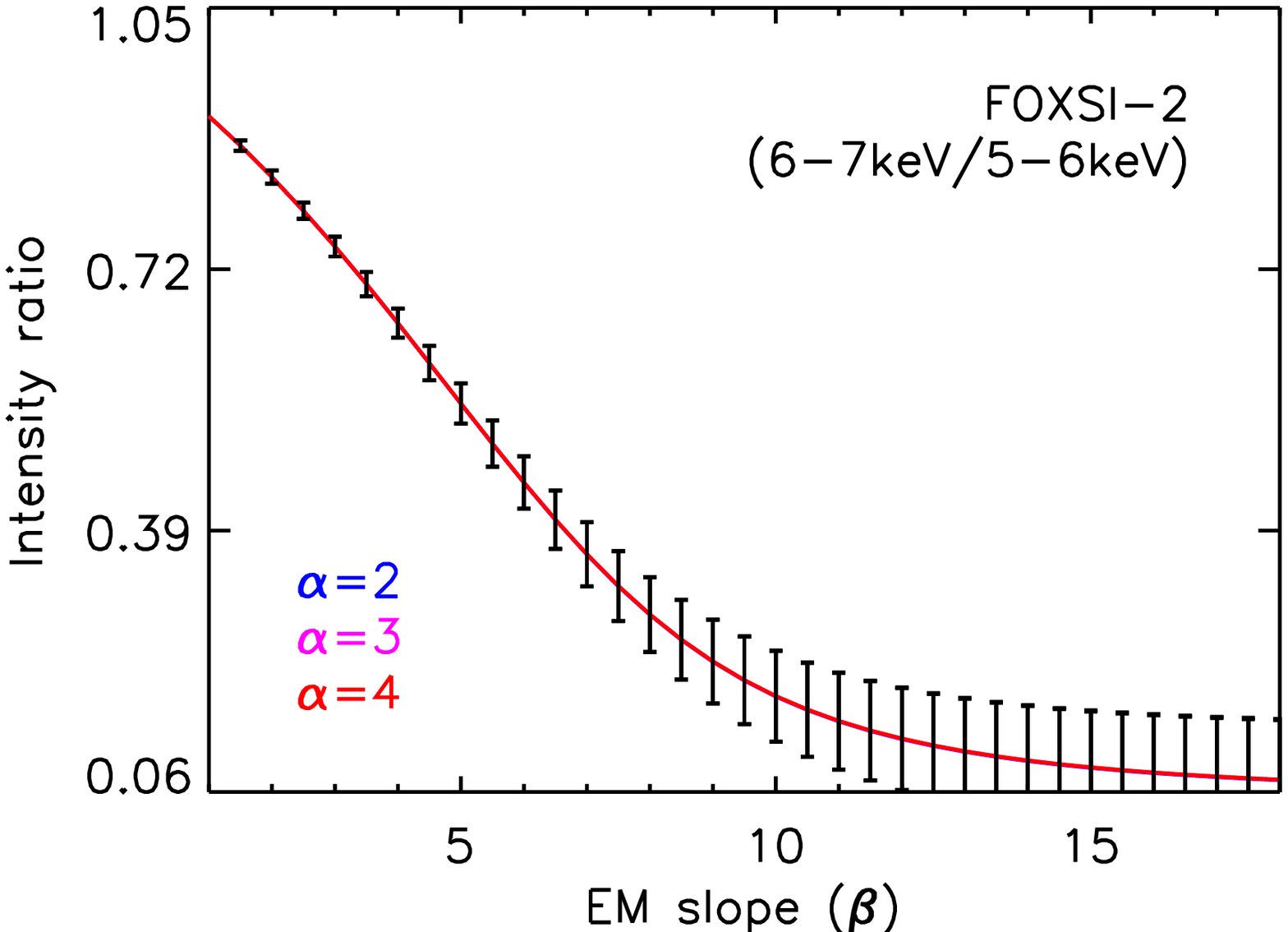}
    \caption{(Left and Middle) The expected intensity in {\foxsi} energy bands 5~-~6 keV and 6~-~7 keV, respectively, as a function of EM slopes. (Right) The ratio of {\foxsi} energy bands as a function  of EM slopes. The expected intensity for EM slopes with different  ${\alpha}$ does not change, indicating that the {\foxsi} intensity is insensitive to   ${\alpha}$. }
    \label{fig:FOXSI_intensity}
  \end{figure}

For this study, we use two of the {\foxsi} energy bands viz., 5~-~6 keV and 6~-~7 keV and consider the temperature response of one of the {\foxsi} telescopes (Athiray et al., to be submitted) to predict the expected intensity for different EM distributions.  We calculate the expected intensity in those two bands using  Equation~\ref{eq:context_eq} and assuming a 30\,s integration time, which was the typical dwell time of the {\foxsi} rocket on a single target.   The left and middle panels in Figure~\ref{fig:FOXSI_intensity} show the predicted intensity in {\foxsi} energy bands with statistical uncertainties for different EM distributions. The intensity ratio of the two {\foxsi} energy bands is shown in the right panel of Figure~\ref{fig:FOXSI_intensity}. We find that {\foxsi} intensity is highly sensitive to   ${\beta}$ and completely insensitive to   ${\alpha}$.




\subsection{{\aia}}

The Atmospheric Imaging Assembly \citep{lemen2012} instrument on the Solar Dynamics Observatory \citep{pesnell_solar_2012} ({\aia}) observes the Sun using different EUV channels sensitive to a range of coronal temperatures.  The EUV channels, created by the selection of multilayer coatings on the optics, were chosen to observe a specific spectral line, but the passbands are broad enough to include multiple spectral lines formed at a variety of temperatures.  For this study, we have considered three {\aia} channels that are sensitive to higher temperature plasma, though each channel also has contributions from lines formed at lower temperature plasma \citep{odwyer2010} that can cause some passbands to have a bi-modal temperature reponse (see, for instance, \citealt{cheung2015}). The 131~{\AA} channel  includes Fe XXI 128.75\,{\AA} line formed at Log $T{\sim}7.05$ and Fe  XXIII 132.91\,\AA\ line formed at Log~$T {\sim}7.15$, but also includes two Fe VIII lines formed at Log $T{\sim}5.6$.   The 94~{\AA} channel includes the Fe XVIII 93.93\,\AA\ line formed at Log $T{\sim}6.85$ and Fe XX 93.78 line formed at Log $T{\sim}7.0$, but also includes the Fe X 94.01 line formed at Log $T{\sim}6.05$. Finally, the strongest line in the 211\,\AA\ channel is Fe XIV 211.31 formed at Log $T {\sim}6.3$.

\begin{figure}[h]
     \includegraphics[width=0.34\linewidth]{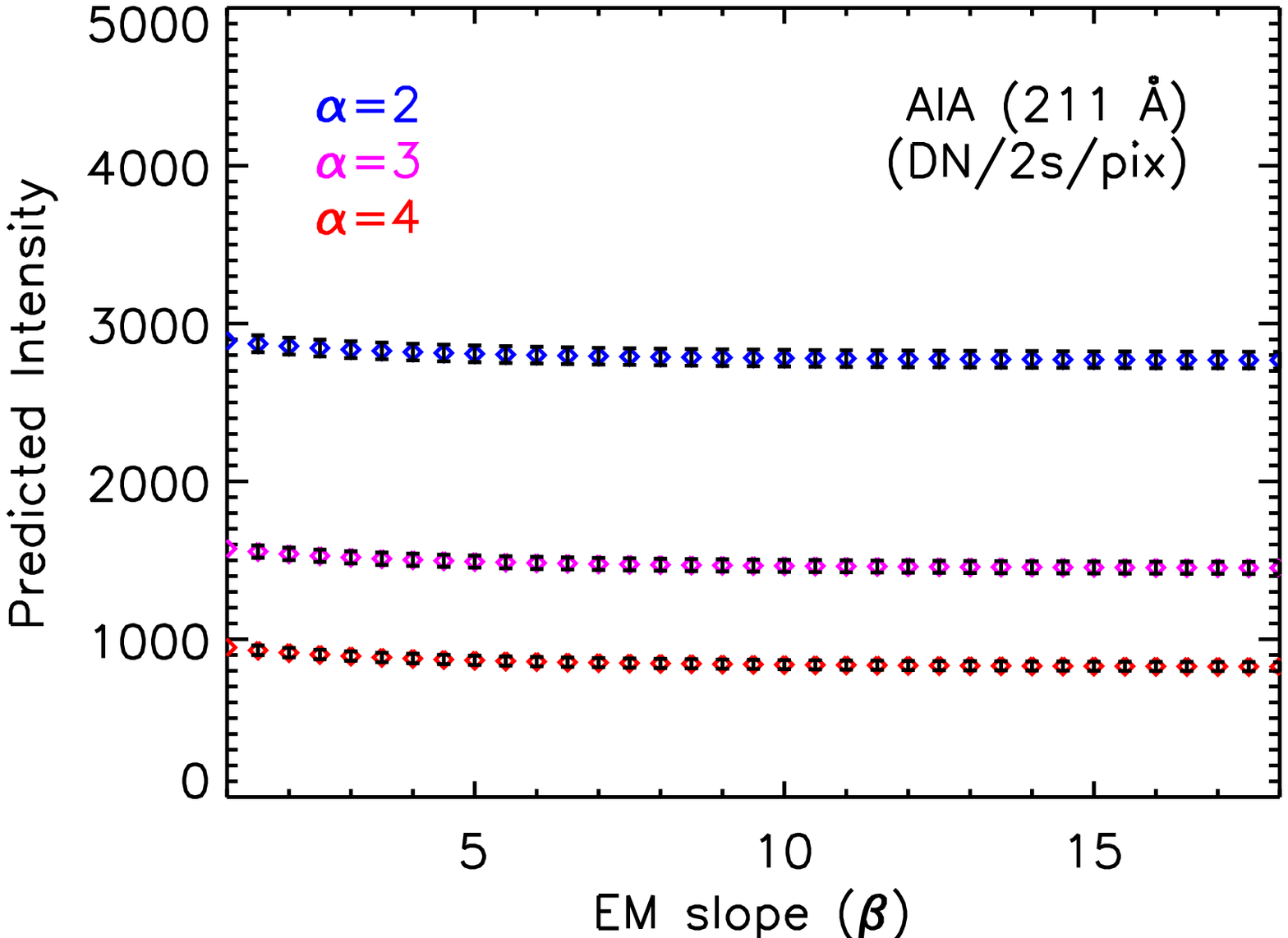}
    \includegraphics[width=0.34\linewidth]{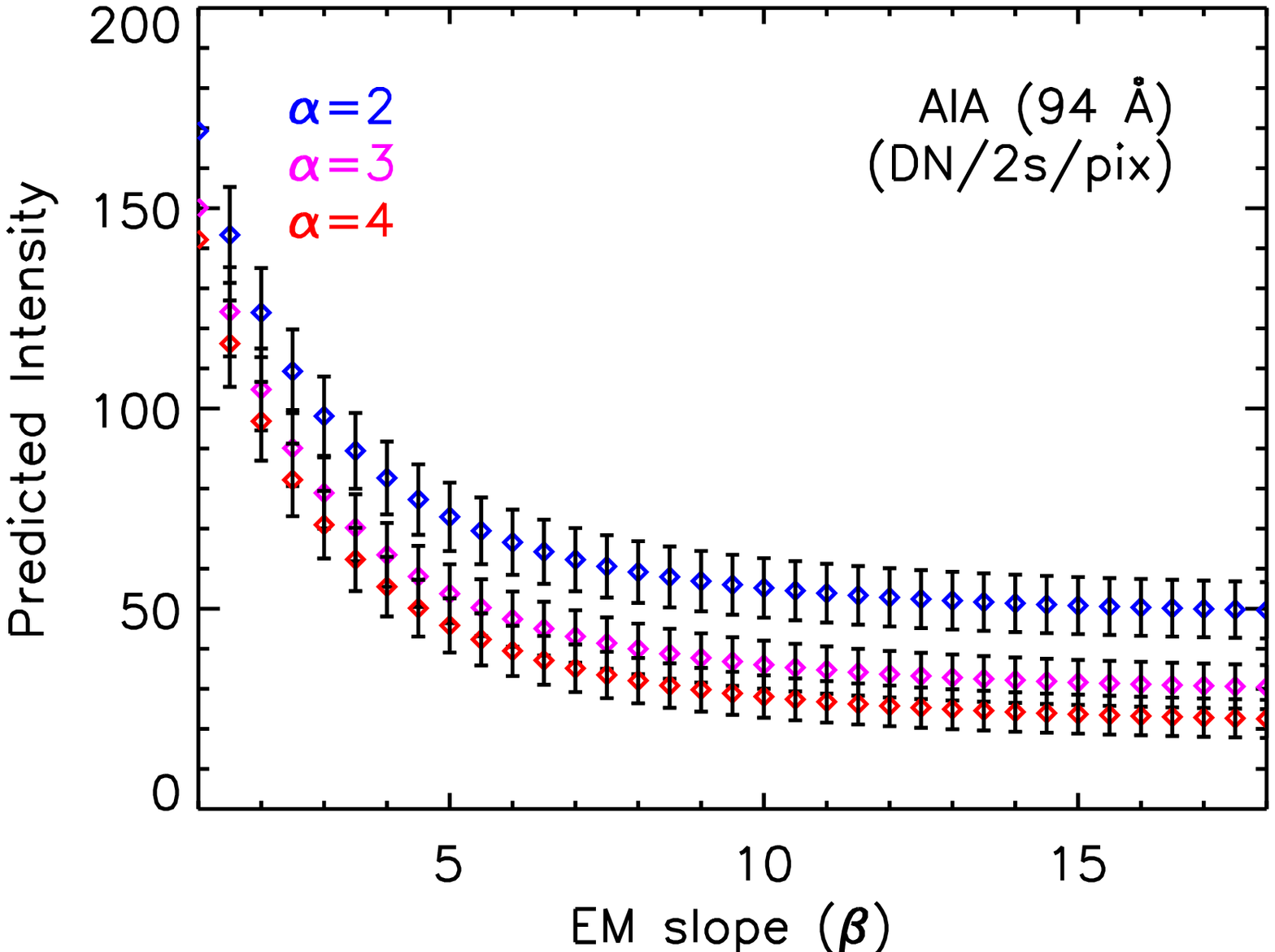}
    \includegraphics[width=0.34\linewidth]{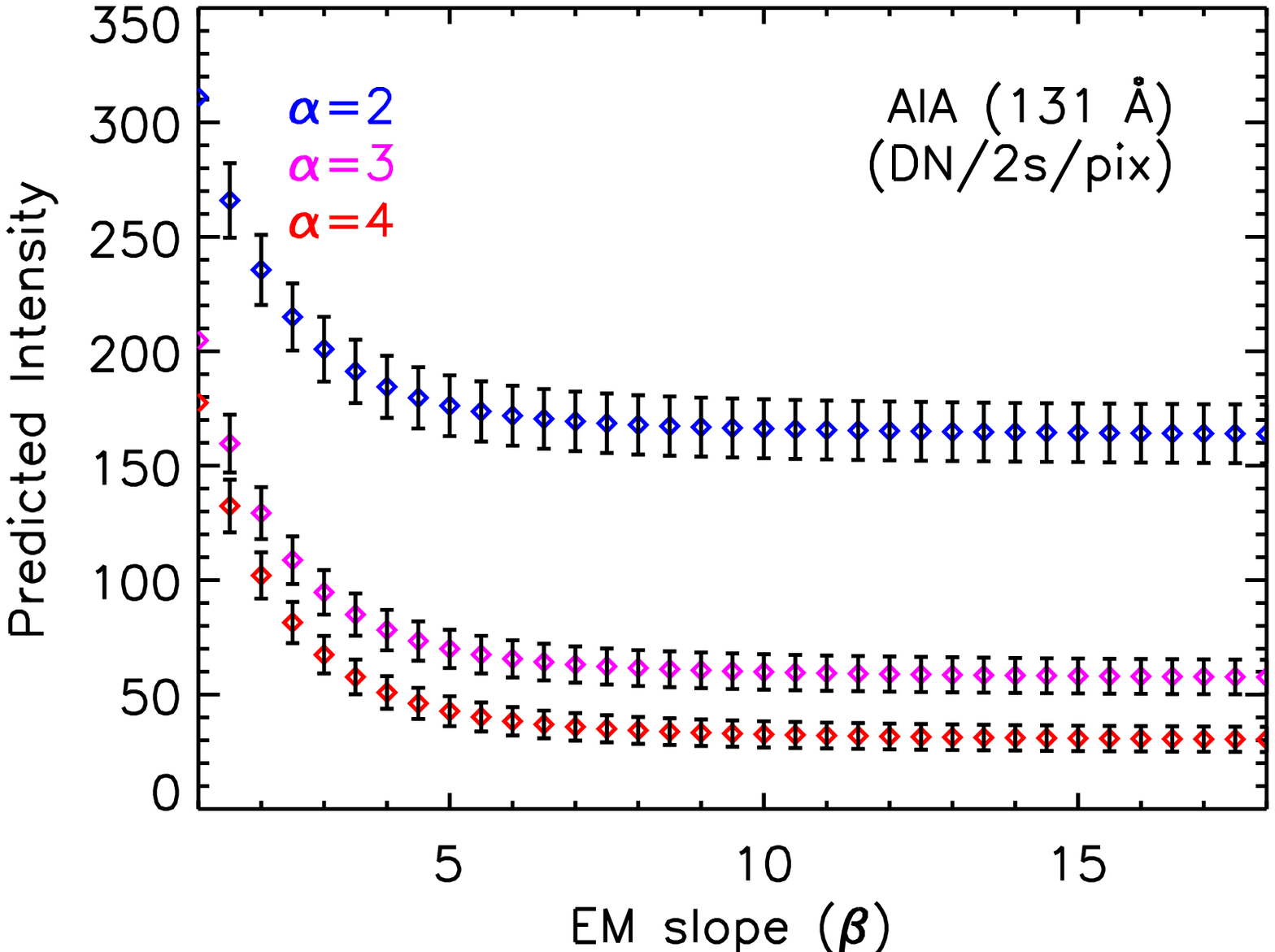}
        \includegraphics[width=0.35\linewidth]{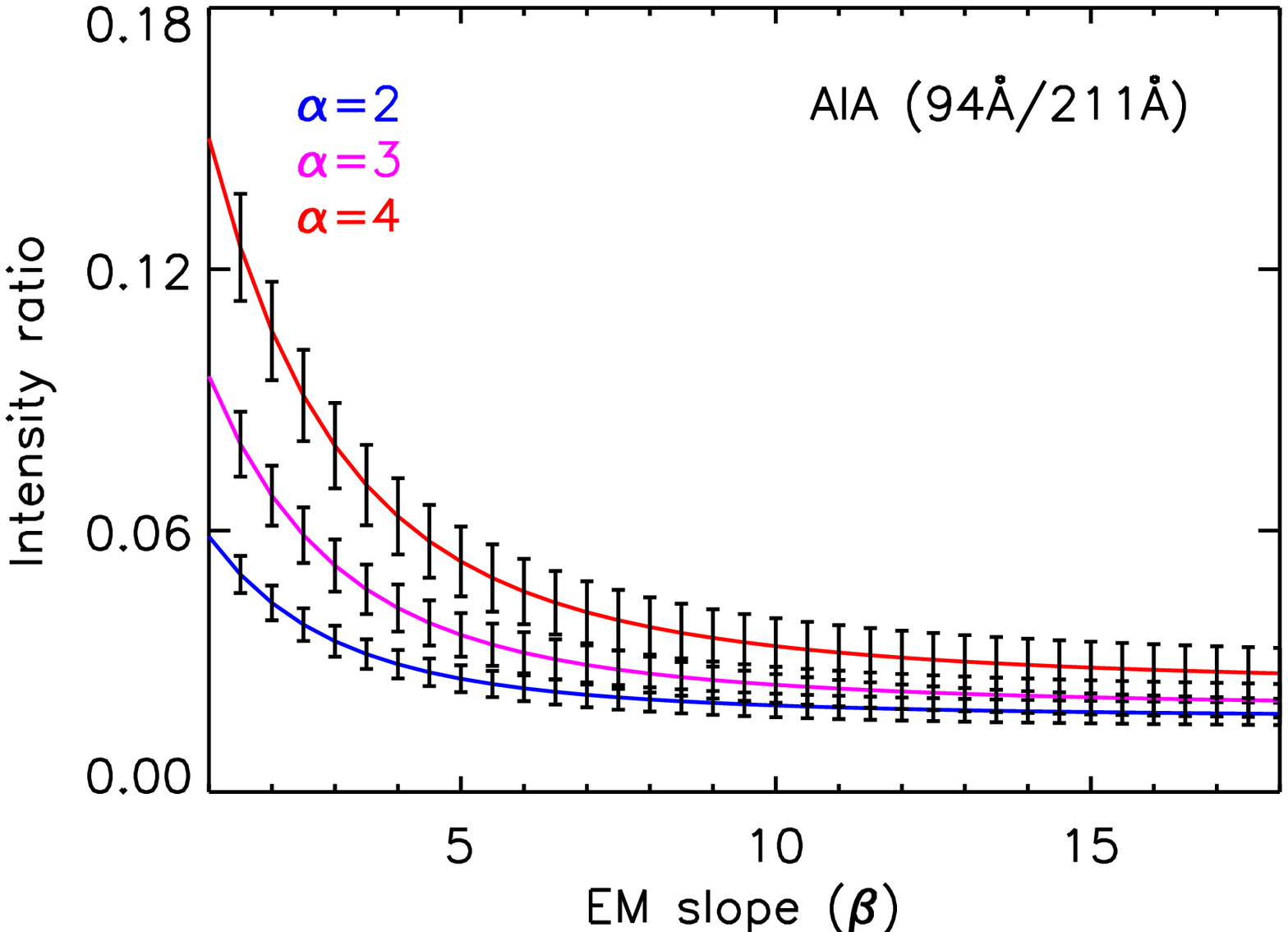}
         \includegraphics[width=0.34\linewidth]{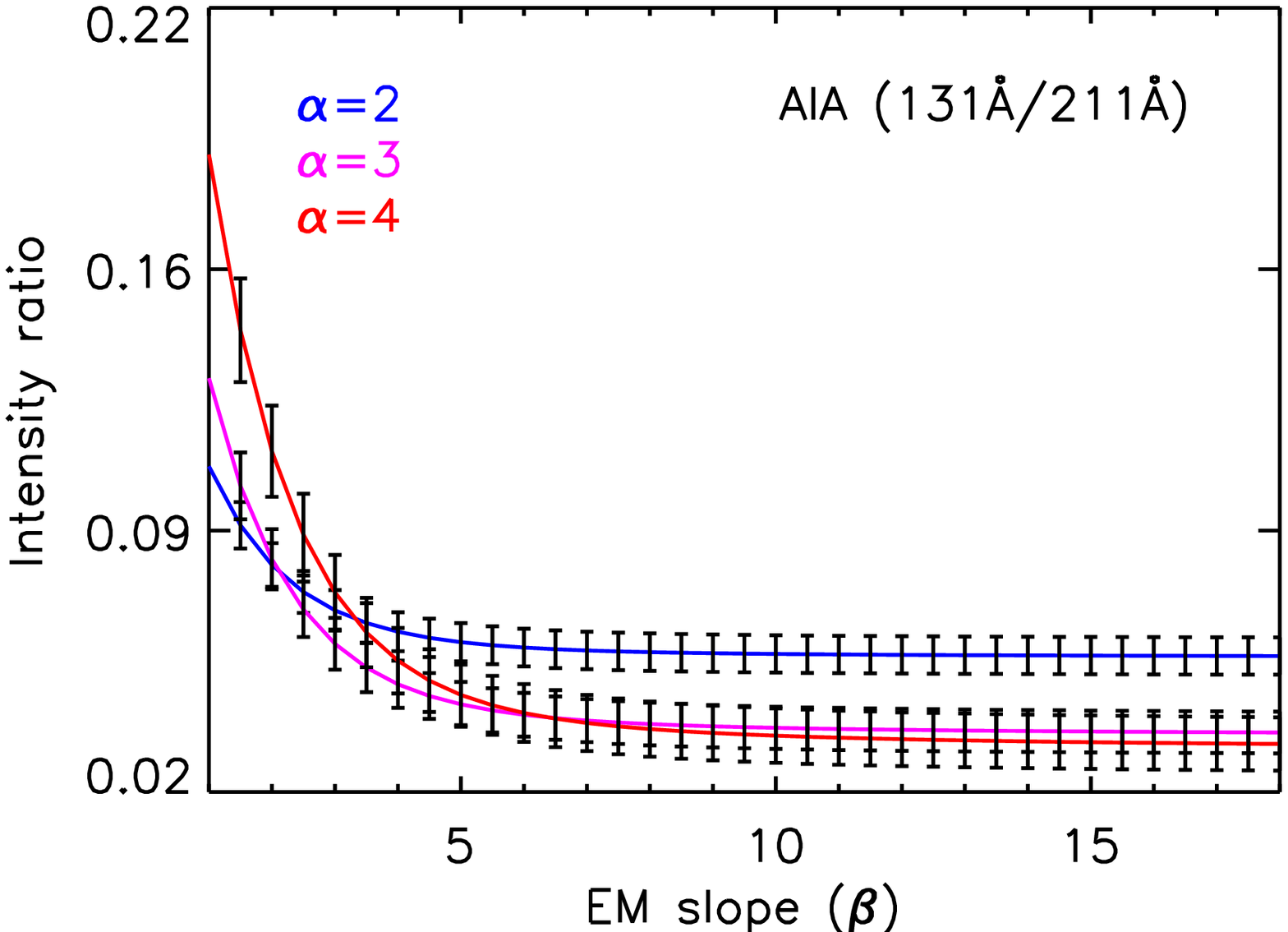}
        \includegraphics[width=0.34\linewidth]{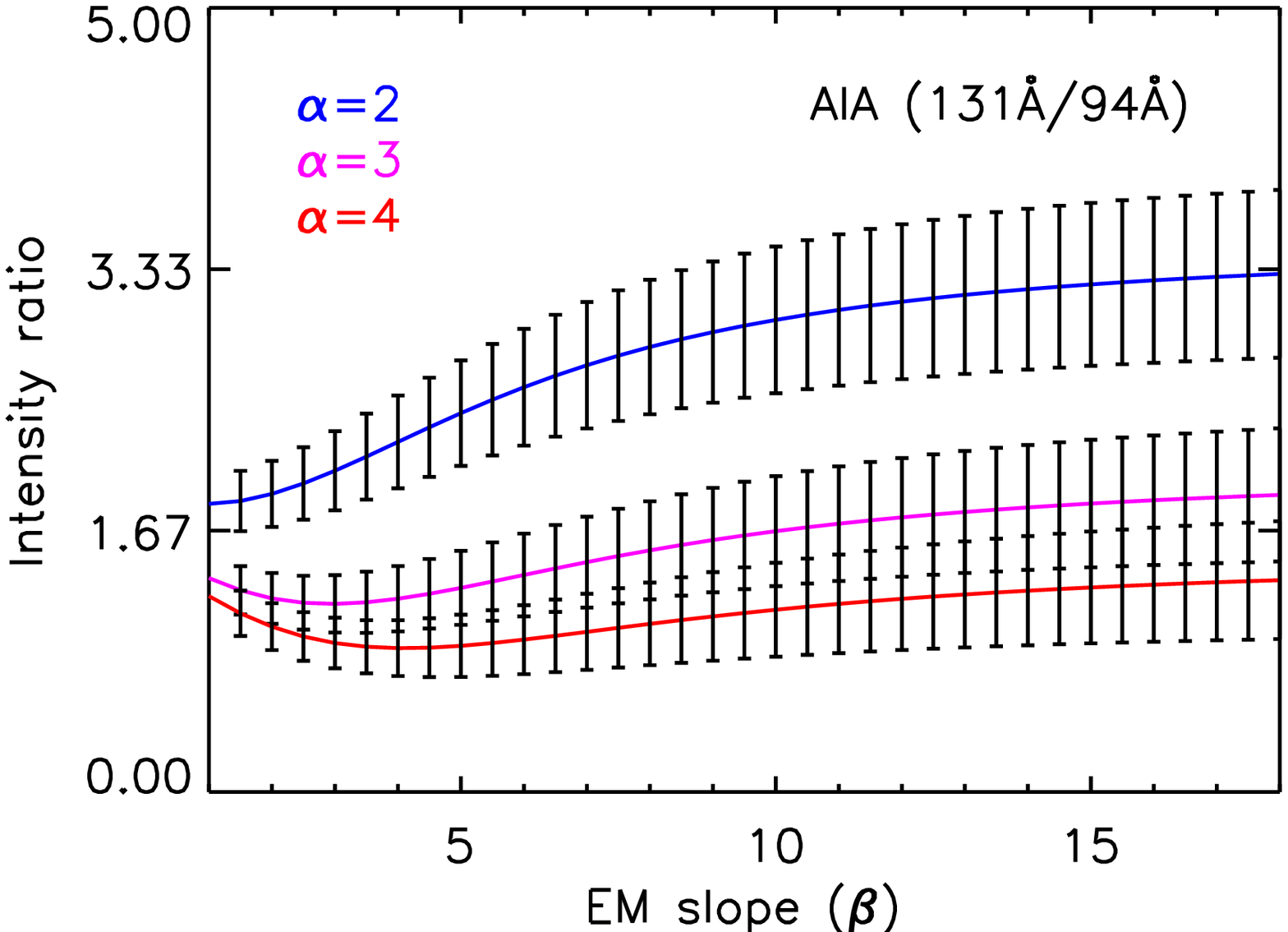}
    \caption{(Top panels): The expected intensity in {\aia} channels 211~\AA, 94~{\AA}, and 131~{\AA} plotted as a function of EM slopes  ${\beta}$. Overplotted in three different colors represent intensity for different  ${\alpha}$'s. (Bottom panels): Intensity  ratios plotted as a function of EM slope ${\beta}$.  The intensity ratios of 94\,\AA\ and 131\,\AA\ to 211\,\AA\ show a weak dependence on ${\beta}$ that varies for different ${\alpha}$, while the 131\,\AA\ to 94\,\AA\ ratio is insensitive ${\beta}$.}
        \label{fig:AIA_flux}
  \end{figure}

To determine whether the ratio of these channel intensities is sensitive to ${\beta}$, we use the temperature response functions from \verb!aia_get_response.pro! with flags \verb!eve_norm! and \verb!chiantifix!. We then calculate the expected intensity in these channels using equation~\ref{eq:context_eq}.  To calculate the error bars, we assume a 2\,s exposure time that is typical for \aia\ images. Plots of expected intensity in the selected \aia\ channels for different EM distributions is shown in the top panels of Figure~\ref{fig:AIA_flux}.   The intensity in 211\,\AA\ channel is insensitive to ${\beta}$ and sensitive to ${\alpha}$, which is expected as the primary lines in the 211\,\AA\ channel is formed at temperatures less than the peak of the emission curve.  The intensity in \aia\ channels 94~{\AA} and 131~{\AA} are sensitive to ${\beta}$, but they are likewise sensitive to   ${\alpha}$, owing to the bi-modal nature of their temperature response.

The intensity ratios plotted as a function of EM slopes is shown in the bottoms panels of Figure~\ref{fig:AIA_flux}. The 94/211\,\AA\ and 131/211\,\AA\ ratios are sensitive to both ${\alpha}$ and ${\beta}$.  The 131/94\,\AA\ ratio appears insensitive to ${\beta}$. We conclude that it would be difficult to accurately determine the high temperature EM slope  ${\beta}$ from only the intensity ratios of \aia\ channels.

\subsection{{\xrt}}

Finally, we consider images from the X-Ray Telescope \citep{golub2007} on board the {\it Hinode} satellite ({\xrt}).   This instrument is a broadband telescope most sensitive to the same soft X-ray wavelength range of \magixs\ (6-25\,\AA).  Different temperature sensitivity is acquired using combinations of different focal plane filters that are available on two  filter wheels.

\begin{figure}[h]
    \includegraphics[width=0.34\linewidth]{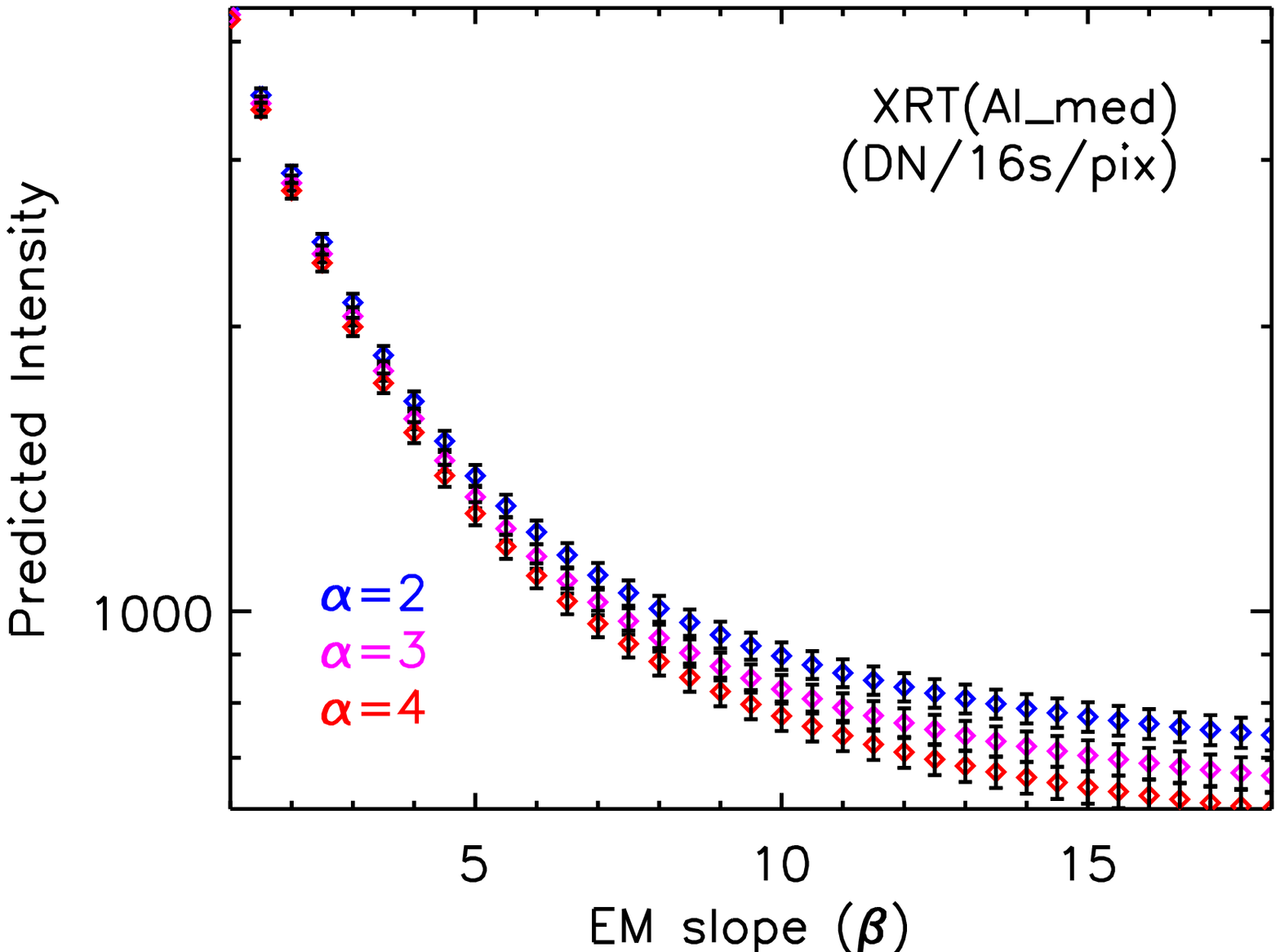}
    \includegraphics[width=0.34\linewidth]{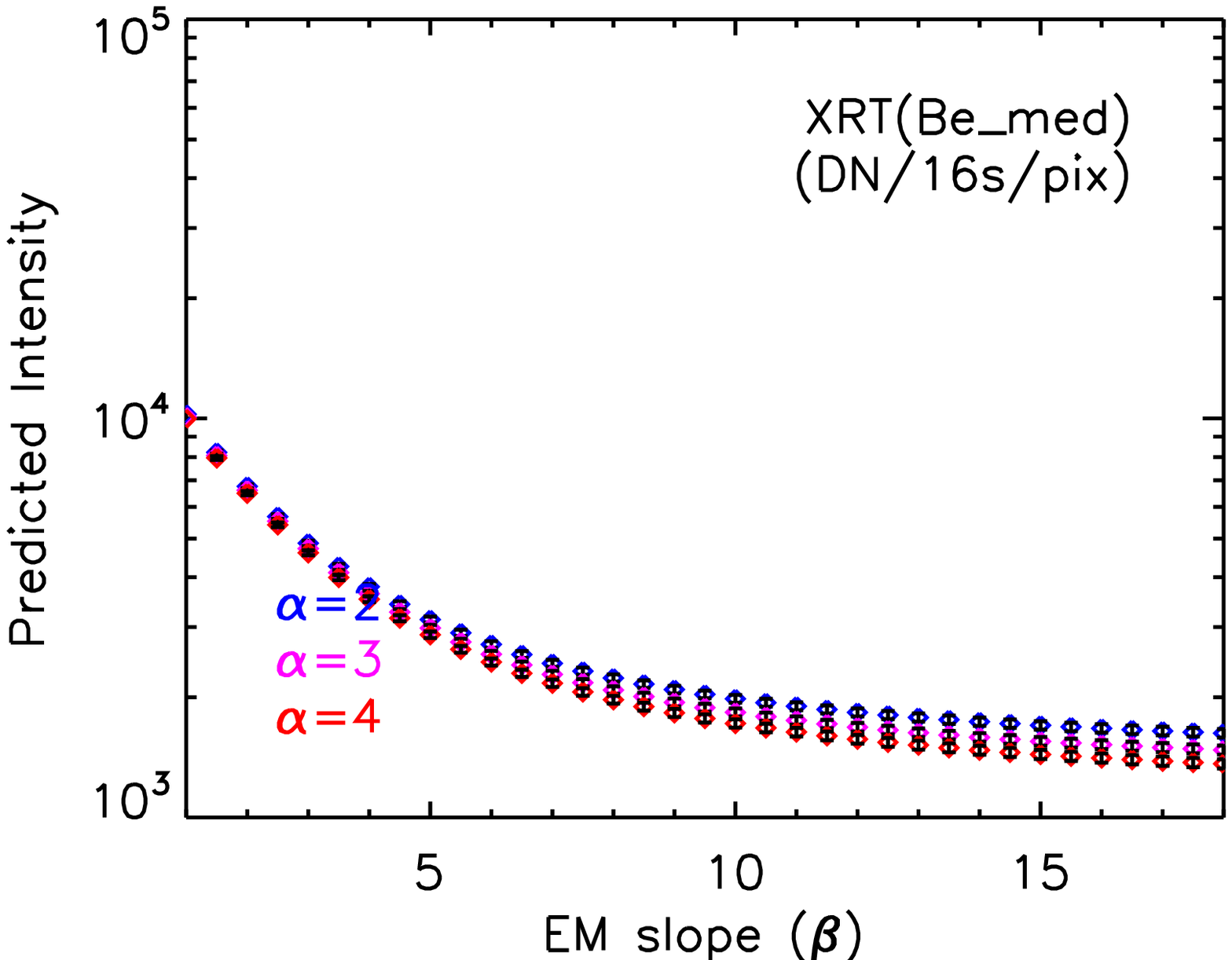}
    \includegraphics[width=0.34\linewidth]{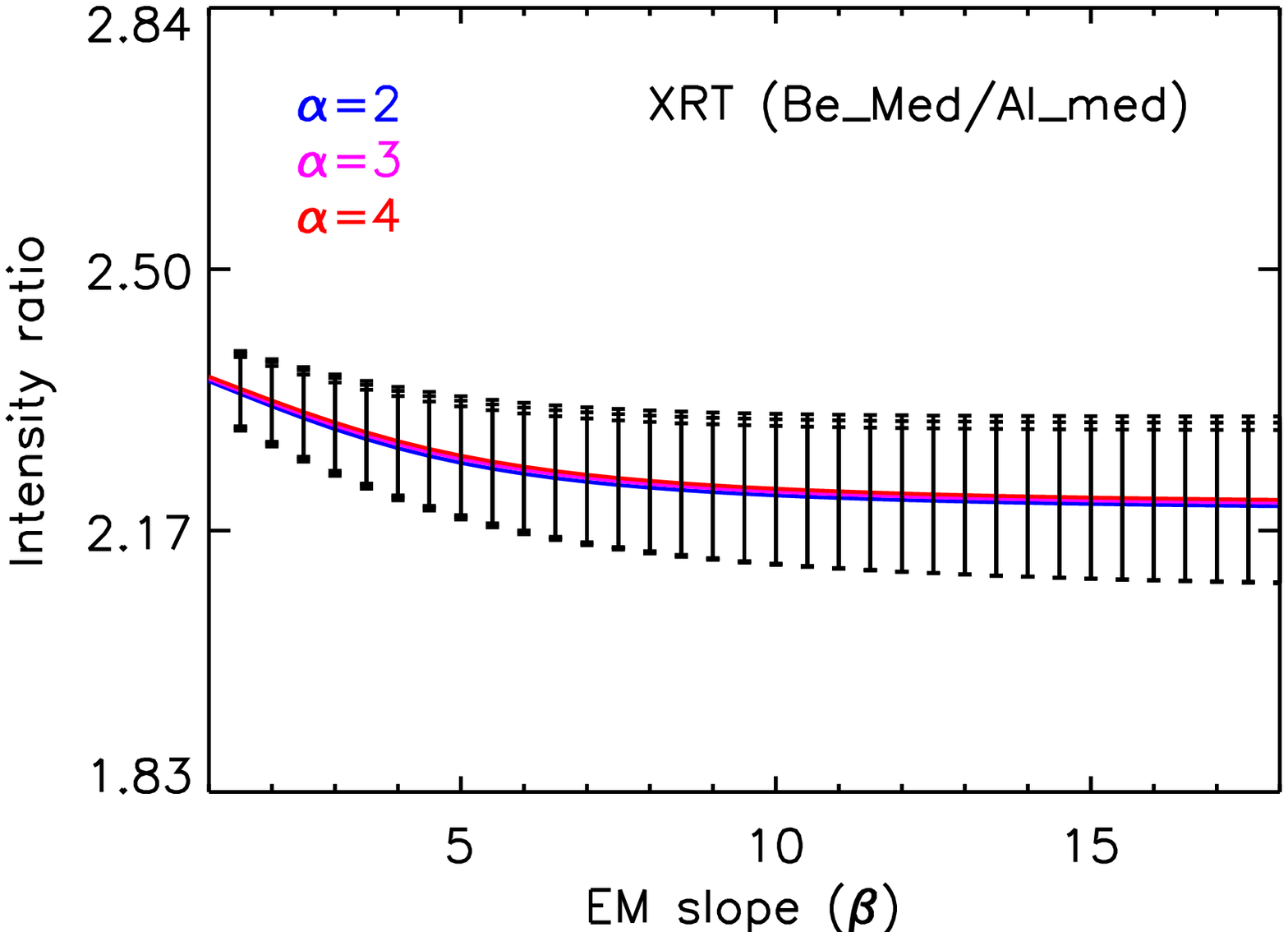}
    \caption{(Left and Middle) The expected intensity in the selected {\xrt} filter combinations plotted as a function of EM slopes. (Right) The ratio of intensity in the two XRT channels plotted as a function of EM slopes. The intensity of XRT channels are sensitive to ${\beta}$ with similar profile. Therefore the intensity ratio of the two XRT channels are insensitive to ${\beta}$.}
    \label{fig:XRT_flux}
  \end{figure}

For this paper, we consider two filter configurations that are useful for the observation of solar active regions in high temperatures viz., Be\_Med/Open and Al\_Med/Open. For the temperature response functions we use \verb!xrt_flux713.pro! (e.g., \citealt{Kobelski2014}) and then  calculate the expected intensity using equation~\ref{eq:context_eq}. To calculate the errors, we assume an exposure time of 16\,s, typical for active region observations in these filters.  The plots of expected intensity in the selected XRT filters for different EM distributions are shown in in the left and middle panels of Figure~\ref{fig:XRT_flux}.  These plots indicate that the considered XRT channels are fairly insensitive to   ${\alpha}$ and sensitive to   ${\beta}$.  The intensity ratio plotted as a function of EM slopes is shown in the right panel of Figure~\ref{fig:XRT_flux}. Though XRT filter intensities are sensitive to   ${\beta}$, they are sensitive to ${\beta}$ in a nearly identical way, meaning the intensity ratios are insensitive to ${\beta}$.

\section{Numerical simulations}

In this section, we use a series of numerical experiments to find the expected line ratios as a function of $\twait$, the time between consecutive heating events on a single strand or the waiting time. We have carried out simulations for the uniform pulse case described in \citet{barnes2016b} using the Enthalpy-Based Thermal Evolution of Loops (EBTEL) model \citep{klimchuk2008,cargill2012}, specifically the two-fluid version of EBTEL as described in \citet{barnes2016a}. We fixed the total simulation time at $8{\times}10^4$ s and varied $\twait$ between 250 s and 5000 s in steps of 250 s for a total of 20 different values of $\twait$. For a given value of $\twait$, the energy per event is the same and because the total simulation time and total input energy are fixed with respect to $\twait$, the energy per event increases with increasing $\twait$. Additionally, we deposit all of the energy into the electrons and assume a symmetric triangular time profile with a fixed duration of 200 s for each heating event. For every simulation, the loop half-length is 40 Mm.

For each EBTEL run, we computed the expected time-dependent intensities in the selected \aia{} channels and {\magixs} lines using the appropriate instrument response (see Section \ref{sec:result}). In the case of the {\magixs} line intensities, we accounted for non-equilibrium ionization when computing the population fractions of Fe 17, 18, and 19. In order to compute the DEM distribution for each value of $\twait$, we applied the regularized inversion method of \citet{Hannah2012} to the time-averaged simulated intensities from each \aia{} channel and {\magixs} line. We chose our temperature bins such that the leftmost edge is at $\log{T}=5.5$ and the rightmost edge at $\log{T}=7.5$ with bin widths of ${\delta}\log T=0.05$. The uncertainties on the \aia{} intensities were obtained using the \verb!aia_bp_estimate_error.pro! routine in SolarSoft \citep[SSW,][]{freeland_data_1998} and we assumed 20\% uncertainty on all of the simulated {\magixs} intensities.

\begin{figure}[h]
    \centering
    \includegraphics[width=0.9\linewidth]{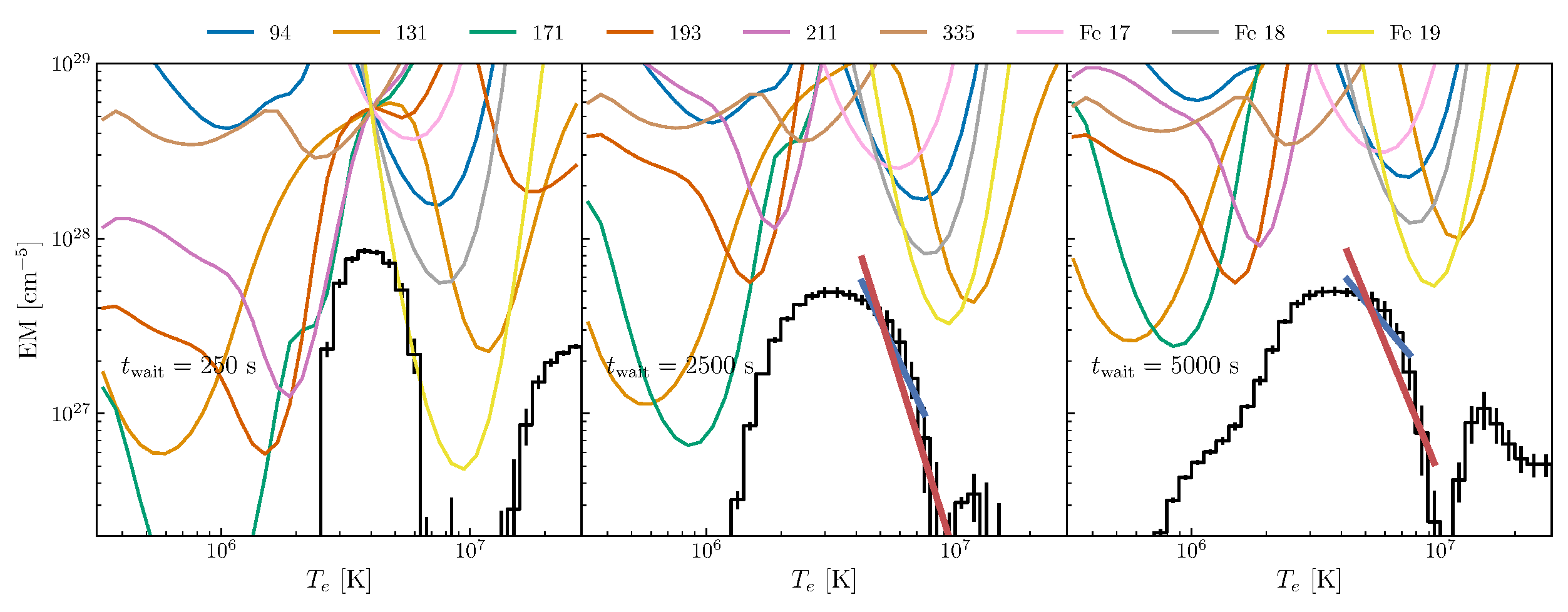}
    \caption{Recovered EM solutions for three different values of $\twait$ using the modeled AIA and {\magixs} intensities. The high-temperature EM slope (${\beta}$) is fit over $6.6{\leq}\log{T}{\leq}6.9$ (blue) and $6.6{\leq}\log{T}{\leq}7.0$ (red).}
    \label{fig:EBTEL_EMsolutions}
  \end{figure}

Figure \ref{fig:EBTEL_EMsolutions} shows a sample of recovered EM solutions for $\twait=250,2500,5000$ s along with the EM loci curves for each of the \aia{} channels and {\magixs} lines. The best fits for $T^{-{\beta}}$ for two temperature ranges, $6.6{\leq}\log{T}{\leq}6.9$ (blue) and $6.6{\leq}\log{T}{\leq}7.0$ (red) are also shown in Figure \ref{fig:EBTEL_EMsolutions}. As expected, we observe broadening of the EM with increasing $\twait$ (i.e. decreasing frequency) as the loop is allowed to evolve over a wider range of both hot and cool temperatures. Most importantly, the amount of emission between 4 MK and 10 MK increases. We also find that including the {\magixs} lines offers a tighter constraint on the hot emission around 10 MK.

  \begin{figure}[h]
    \centering
    \includegraphics[width=0.3\linewidth]{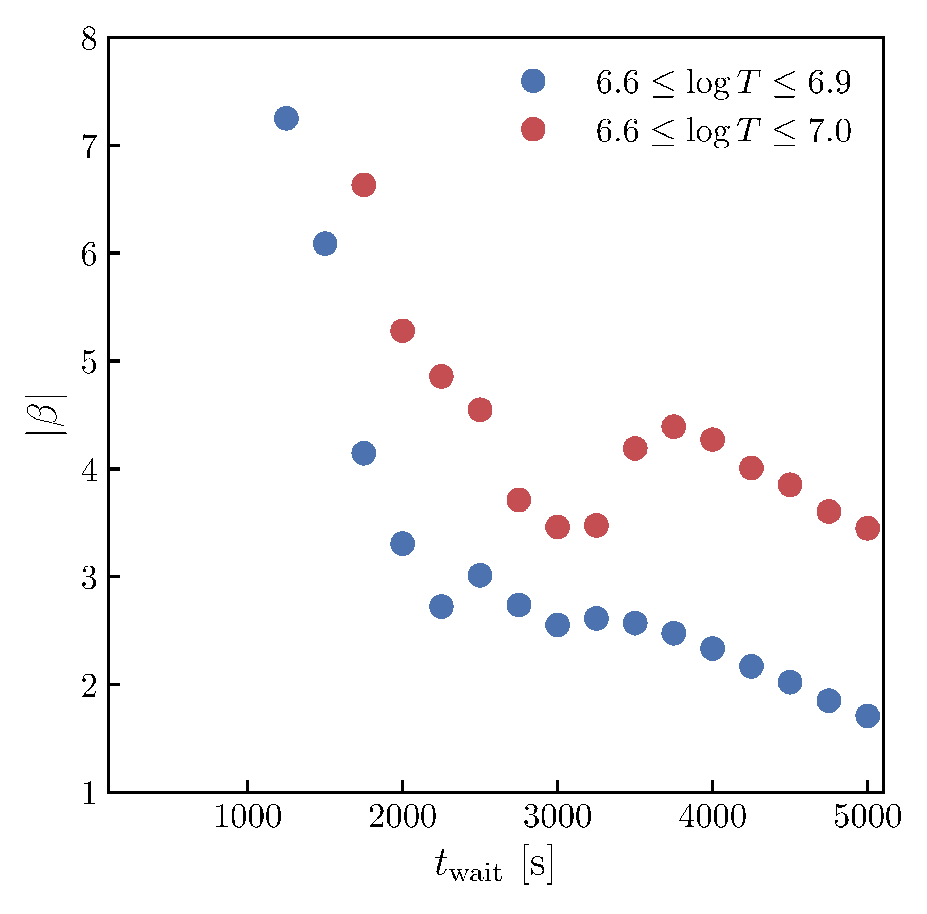}
    \caption{Dependence of the high-temperature EM slope, ${\beta}$, on the waiting time between heating events, $\twait$, obtained from the EBTEL simulations and EM inversion.}
    \label{fig:EBTEL_beta}
  \end{figure}
\begin{figure}[h]
    \centering
     \includegraphics[width=0.75\linewidth]{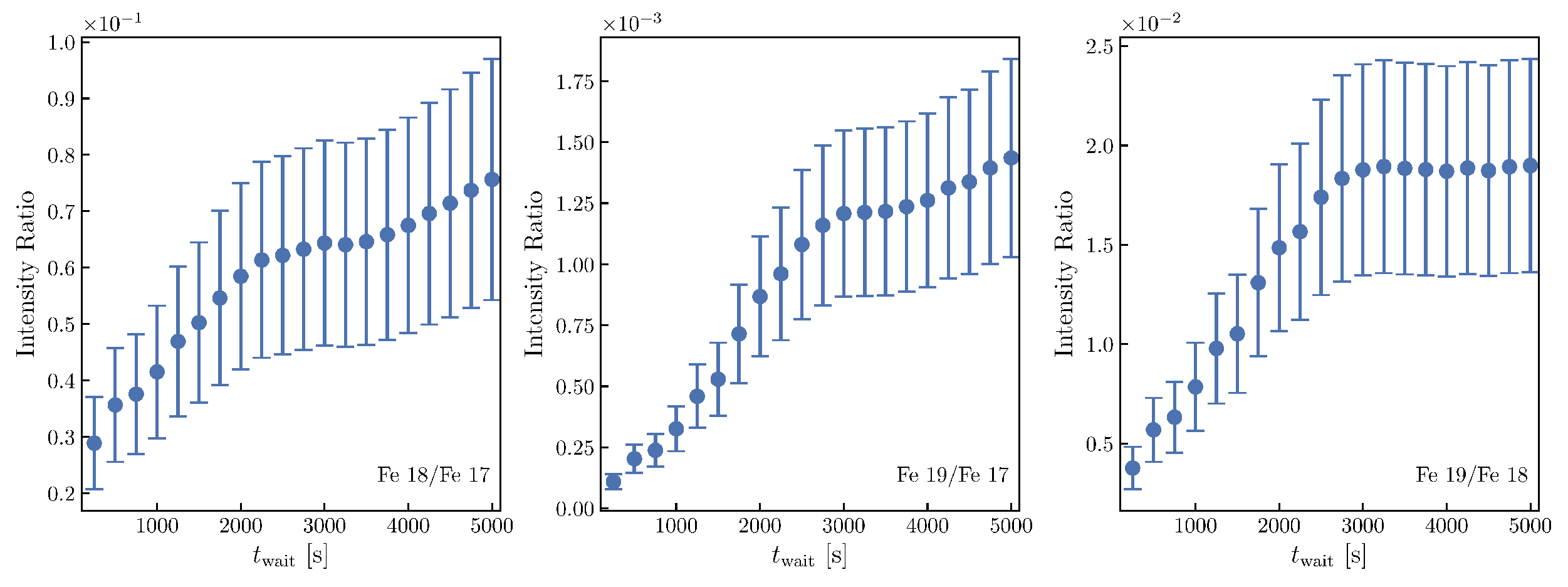}
    \caption{Intensity ratio as a function of $\twait$ for the three possible combinations of {\magixs} lines: Fe 18/17 (left), Fe 19/17 (middle), and Fe 19/18 (right). The error bars on each intensity ratio were calculated by adding the uncertainties on each intensity in quadrature.}
    \label{fig:EBTEL_ratios}
  \end{figure}
  Figure~\ref{fig:EBTEL_beta} shows ${\beta}$, as computed from the fits shown in Figure~\ref{fig:EBTEL_EMsolutions}, for each value of $\twait$. The different colors denote the different temperature ranges over which the fit was computed. If T$^{-{\beta}}$ could not be fit over the given temperature interval for a particular value of $\twait$, that point is not included in the plot. As the waiting time between heating events increases, ${\beta}$ decreases, consistent with the increase in hot emission expected from low-frequency heating.  As deriving ${\beta}$ is sensitive to the range of temperature considered (see Figure~\ref{fig:EBTEL_beta}), same temperature range as in the analysis should be used to determine the heating frequency. The inversion feature at $\twait{\sim}3250$ s for the extended temperature interval (red) is due to emission at T $>$ 10 MK in the recovered EM distribution. This excess hot emission is likely an artifact of the EM reconstruction, since it is associated with very low emission measure at temperatures where the distribution is poorly constrained by the contribution functions. This feature is not present in the values of ${\beta}$ computed from the shorter fit interval (blue) as that interval does not span those temperatures at which the EM is poorly constrained. Finally, Figure \ref{fig:EBTEL_ratios} shows the intensity ratios as a function of $\twait$ for the three possible combinations of {\magixs} lines. The uncertainty of each ratio, ${\sigma}_r$, is calculated by adding the uncertainties of each intensity ($0.2I$) in quadrature such that ${\sigma}_r=0.2r\sqrt{2}{\approx}0.3r$, where $r$ is the intensity ratio. Despite the relatively large error bars, each ratio still shows a clear dependence on $\twait$.  This confirms the intensity ratios themselves can be used as a diagnostic for the heating frequency, even in the case where the entire emission measure distribution cannot be calculated. Our results strongly motivate the idea of using MaGIXS line intensity ratios to determine the heating frequency and, is better than deriving just ${\beta}$, which is very sensitive to the range of temperature considered.

\section{Discussion}

The high temperature component of the emission measure in different solar structures is thought to be a ``smoking gun'' observation needed to limit the frequency of solar heating events. In this paper, we investigated the sensitivity of intensity ratios available on current or future instruments to the high temperature component of the emission measure. We assumed broken power-law EM distributions with different slopes ${\alpha}$ (Log T = 6.0 to 6.6) and ${\beta}$ (Log T = 6.6 to 7.1) (see Figure~\ref{fig:EMslopes}).

We first considered the new instrument, \magixs, being developed for a sounding rocket flight in April 2020.  Our analysis shows the line ratios of some of the strongest lines that {\magixs} will observe are sensitive to ${\beta}$ and insensitive to ${\alpha}$. Therefore, the ratio between any two of the line intensity from {\magixs} can be used as a proxy to determine the high temperature EM slope ${\beta}$. Further we also showed that the uncertainty in the high temperature emission measure slope (${\sigma}_{\beta}/{\beta}$) will be more tightly constrained by \magixs\ than it has been in previous investigations \citep[e.g.,][]{warren2012}.  Interpretation of the \magixs\ line ratios, and relating these ratios to the underlying heating information, however requires additional numerical modeling, similar to \cite{barnes2016a,barnes2016b}.  One aspect of using spectral line ratios to determine the impulsive heating parameters is the ionization timescales, meaning the emitting plasma may not be in ionization equilibrium as assumed here. The timescale of ionization depends strongly on the density of the heated plasma, which varies strongly with heating parameters, hence is not possible to determine its impact without detailed modeling.

In addition to the broken power-law model, we also performed a set of numerical experiments using EBTEL for 20 different values of $\twait$ ranging from high- to low-frequency heating. Using the simulated {\aia} and {\magixs} intensities, we computed the EM distribution through the regularized inversion method of \citet{Hannah2012} and used the resulting EM to calculate ${\beta}$ and the {\magixs} intensity ratios as a function of $\twait$. We found that ${\beta}$ decreased with increasing $\twait$ (decreasing frequency) though the value of ${\beta}$ was very sensitive to the limits over which $T^{-{\beta}}$ was fit to the EM distribution. Additionally, we found that all three intensity ratios were significantly sensitive to $\twait$ such that they could be used as a diagnostic for the heating frequency. Results from our investigation suggest that {\magixs} line intensity ratios are better diagnostic to determine heating frequency than deriving ${\beta}$, which is very sensitive to the range of temperature considered.

\begin{figure}[h]
    \includegraphics[width=0.5\linewidth]{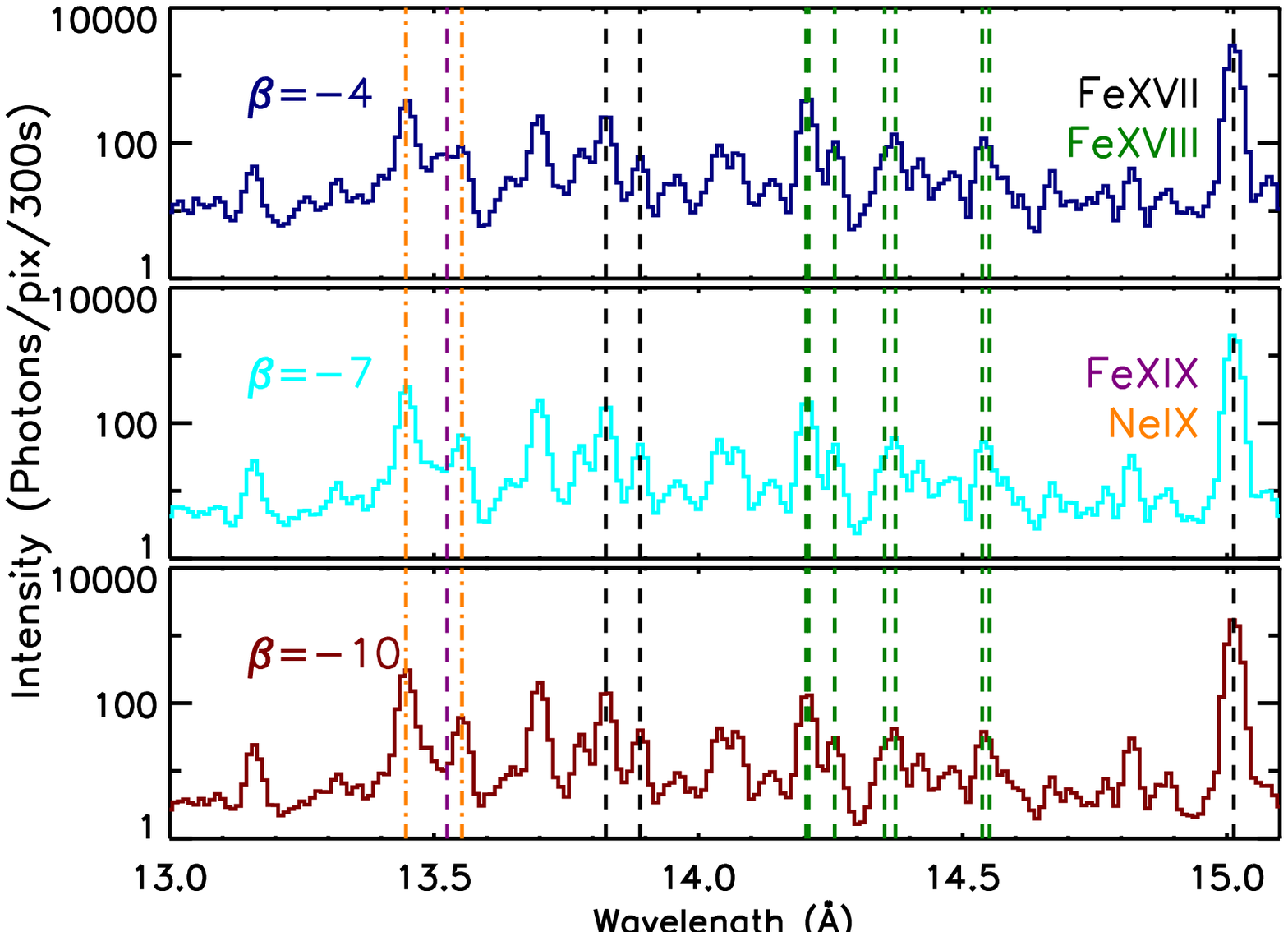}
    \includegraphics[width=0.5\linewidth]{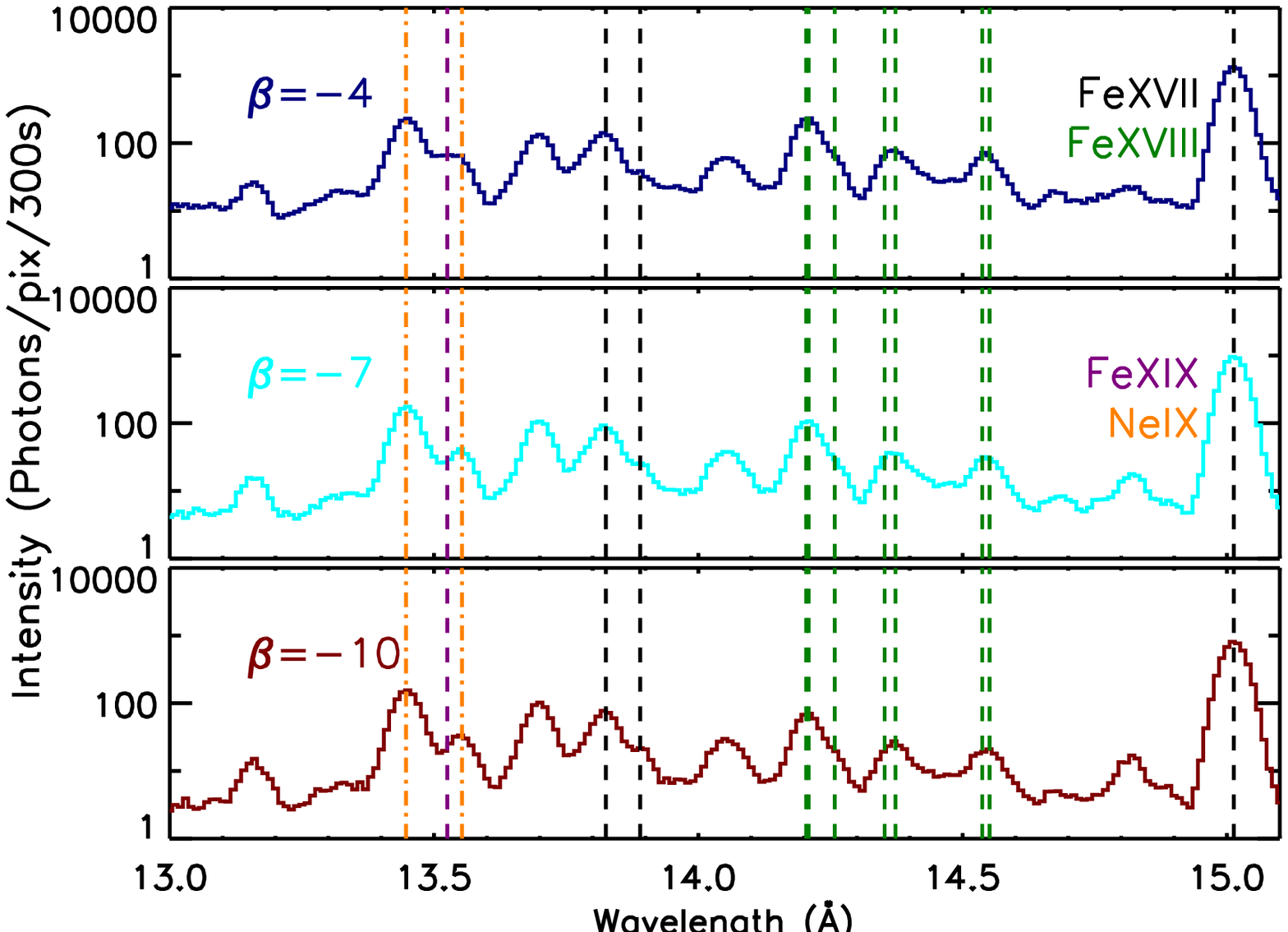}
    \caption{ The predicted {\magixs}  spectrum for three different EM distributions with ${\alpha}$=3 and ${\beta}$ = -4, -7, -10 is shown in three different colors. The dashed vertical lines indicate some of the strong lines of Fe XVII, Fe XVIII, and  Fe XIX that {\magixs} will observe. The dashed orange lines indicate Ne IX formed at Log T$_{max}$=6.6, which is very close to the strong Fe XIX line. The spectrum shown here includes an instrumental broadening of FHWM = 0.022 {\AA} (Left) and 0.05 {\AA}(Right), which is the worst case, and contribution from Poisson deviates random noise.}
    \label{fig:magixspredictedspectra}
\end{figure}

For this analysis, we selected the three brightest spectral lines of Fe XVII, Fe XVIII, and Fe XIX. Using the instrument response, we have calculated the expected {\magixs} intensity summed over the entire line for a series of assumed EM distributions with range of EM slopes ${\alpha}$ and ${\beta}$. We have not considered potential of line blends.  Figure~\ref{fig:magixspredictedspectra} shows the expected signal in the portion of the \magixs\ spectrum including these three spectra lines.  For this figure, we have included an instrumental broadening of 0.05\,\AA, which is the instrument requirement for \magixs.  The actual instrumental broadening will be measured prior to flight.  This figure shows that Fe XVII and Fe XVIII are well isolated lines and are blended only with spectral lines emitted by the same ions.  Fe XIX, however, is blended with a pair of Ne IX lines, formed at Log $T{\sim}6.6$.  We are considering methods of removing this blend and determining the Fe XIX line intensity.  Additionally, we have used this analysis to derive the calibration requirements for the \magixs\ instrument.  We require that the relative uncertainty in the \magixs\ instrument response function contribute less than 10\% to the uncertainty of ${\beta}$.  Finally, line ratios such as the ones discussed here for \magixs\ may be useful to limit the coronal heating parameters (see \citealt{barnes2016b}).   However, we also plan to complete full emission measure calculations with the \magixs\ spectra both alone and with other complementary observations.

Similarly, we demonstrated the ratio of intensities in different \foxsi\ energy ranges is also sensitive to the high temperature slope and insensitive to the cool temperature slope, ${\alpha}$.  \foxsi\ has the added benefit of having the majority of its temperature sensitivity originating from bremsstrahlung emission and, hence, would not be impacted by ionization timescales.  However, the spatial resolution of \foxsi\ (${\sim}$ 30\arcsec, half power diameter) limits the ability to resolve individual structures in active regions.  A combination of \magixs\ and \foxsi\ would be a unique data set that could be used to probe coronal heating in a variety of structures and provide additional information on the ionization state of the emitting plasma.

We also considered intensity  ratios for other, existing instruments that are commonly used to probe the hot plasma of the solar corona. We found that it is not easy to determine ${\beta}$ from the intensity of {\aia} channels. Though the \aia\ 335\,\AA\ channel, formed at $\log{T} {\sim}6.4$, could be useful for this type of study, we do not discuss the results here due to a drop in temperature sensitivity from contamination accumulation for this channel \citep{Boerner2014}. However, the ratios including 335\,\AA\ are not strongly sensitive to ${\beta}$ and do not change the conclusions derived from this work.

We found that ${\beta}$ could not be easily determined from the intensity ratios of the {\xrt} channels. This once again confirms the blind-spot demonstrated in \citet{winebarger2012}; namely that there is limited sensitivity of existing instrumentation to plasma at temperature $>$ 5 MK. The reason could be due to the XRT filter response which is fairly broad and the relative variations in the response do not have the resolution to separate high temperature emission.  Results from \citet{Ishikawa2019} also confirm that plasma below 5 MK chiefly contribute to the total flux observed in some of the XRT filters. Therefore the intensity ratios of XRT channels would not directly yield the high temperature EM slope. Recent results from \citet{Ishikawa2019} clearly show that using the Reuven Ramaty High Energy Solar Spectroscopic Imager ({\rhessi}) HXR data between 3~-~8 keV  provides a well-constrained high temperature slope below 8 MK. Though {\rhessi} was sensitive to high temperature plasma emission, it had limited instrument sensitivity for small active regions due to indirect imaging and high detector background.

In this paper, we chose a peak emission measure value of $8 {\times}10^{27}$ cm$^{-5}$, consistent with an average size active region structure in \cite{warren2012}.  This value ultimately determines the counts that are simulated in each instrument, and, in turn, determines the level of uncertainty in the line ratios.  If the structures imaged by any of these instruments had a higher or lower emission measure value, the uncertainty in the line ratios and the resulting ${\beta}$ would be lower or higher, respectively.

P.S. Athiray`s research is supported by an appointment to the NASA Postdoctoral Program at the Marshall Space Flight Center, administrated by Universities Space Research Association under contract with NASA. Harry P. Warren was funded by NASA's Hinode program. The {\magixs} instrument team is supported by NASA Low Cost Access to Space program.

\software{
    Astropy \citep{the_astropy_collaboration_astropy_2018},
    IPython \citep{perez_ipython_2007},
    matplotlib \citep{hunter_matplotlib_2007},
    numpy \citep{oliphant_guide_2006},
    scipy \citep{jones_scipy_2001},
    seaborn \citep{waskom_seaborn_2018},
    SSW \citep{freeland_data_1998}
}

\bibliographystyle{aasjournal}
\bibliography{references,solar}

\begin{thebibliography}{}
\expandafter\ifx\csname natexlab\endcsname\relax\def\natexlab#1{#1}\fi

\bibitem[{{Asgari-Targhi} \& {van Ballegooijen}(2012)}]{asgaritarghi2012}
{Asgari-Targhi}, M., \& {van Ballegooijen}, A.~A. 2012, \apj, 746, 81

\bibitem[{{Athiray} {et~al.}(2017){Athiray}, {Buitrago-Casas}, {Bergstedt},
  {Vievering}, {Musset}, {Ishikawa}, {Glesener}, {Takahashi}, {Watanabe},
  {Courtade}, {Christe}, {Krucker}, {Goetz}, \& {Monson}}]{Athiray2017}
{Athiray}, P.~S., {Buitrago-Casas}, J.~C., {Bergstedt}, K., {et~al.} 2017, in
  Society of Photo-Optical Instrumentation Engineers (SPIE) Conference Series,
  Vol. 10397, Society of Photo-Optical Instrumentation Engineers (SPIE)
  Conference Series, 103970A

\bibitem[{{Barnes} {et~al.}(2016{\natexlab{a}}){Barnes}, {Cargill}, \&
  {Bradshaw}}]{barnes2016a}
{Barnes}, W.~T., {Cargill}, P.~J., \& {Bradshaw}, S.~J. 2016{\natexlab{a}},
  \apj, 829, 31

\bibitem[{{Barnes} {et~al.}(2016{\natexlab{b}}){Barnes}, {Cargill}, \&
  {Bradshaw}}]{barnes2016b}
---. 2016{\natexlab{b}}, \apj, 833, 217

\bibitem[{{Boerner} {et~al.}(2014){Boerner}, {Testa}, {Warren}, {Weber}, \&
  {Schrijver}}]{Boerner2014}
{Boerner}, P.~F., {Testa}, P., {Warren}, H., {Weber}, M.~A., \& {Schrijver},
  C.~J. 2014, \solphys, 289, 2377

\bibitem[{{Bradshaw} {et~al.}(2012){Bradshaw}, {Klimchuk}, \&
  {Reep}}]{bradshaw2012}
{Bradshaw}, S.~J., {Klimchuk}, J.~A., \& {Reep}, J.~W. 2012, \apj, 758, 53

\bibitem[{{Brooks} {et~al.}(2012){Brooks}, {Warren}, \&
  {Ugarte-Urra}}]{brooks2012}
{Brooks}, D.~H., {Warren}, H.~P., \& {Ugarte-Urra}, I. 2012, \apjl, 755, L33

\bibitem[{{Brooks} {et~al.}(2013){Brooks}, {Warren}, {Ugarte-Urra}, \&
  {Winebarger}}]{brooks2013}
{Brooks}, D.~H., {Warren}, H.~P., {Ugarte-Urra}, I., \& {Winebarger}, A.~R.
  2013, \apjl, 772, L19

\bibitem[{{Brosius} {et~al.}(2014){Brosius}, {Daw}, \& {Rabin}}]{brosius2014}
{Brosius}, J.~W., {Daw}, A.~N., \& {Rabin}, D.~M. 2014, \apj, 790, 112

\bibitem[{{Cargill}(2014)}]{cargill2014}
{Cargill}, P.~J. 2014, \apj, 784, 49

\bibitem[{{Cargill} {et~al.}(2012){Cargill}, {Bradshaw}, \&
  {Klimchuk}}]{cargill2012}
{Cargill}, P.~J., {Bradshaw}, S.~J., \& {Klimchuk}, J.~A. 2012, \apj, 752, 161

\bibitem[{{Champey} {et~al.}(2016){Champey}, {Winebarger}, {Kobayashi},
  {Savage}, {Cirtain}, {Cheimets}, {Hertz}, {Golub}, {Ramsey}, {McCracken},
  {Marquez}, {Allured}, {Heilmann}, {Schattenburg}, \&
  {Bruccoleri}}]{doi:10.1117/12.2232820}
{Champey}, P., {Winebarger}, A., {Kobayashi}, K., {et~al.} 2016, in \procspie,
  Vol. 9905, Space Telescopes and Instrumentation 2016: Ultraviolet to Gamma
  Ray, 990573

\bibitem[{{Champey} {et~al.}(2019){Champey}, {Athiray}, {Winebarger},
  {Kobayashi}, , {Savage}, , {Cheimets}, {Hertz}, {Ramsey}, {Ranganathan},
  {Marquez}, {Allured}, {Parker}, {Heilmann}, \& {Schattenburg}}]{champey2019}
{Champey}, P., {Athiray}, P.~S., {Winebarger}, A.~R., {et~al.} 2019, in Society
  of Photo-Optical Instrumentation Engineers (SPIE) Conference Series, Vol.
  11119, Optics for EUV, X-Ray, and Gamma-Ray Astronomy IX, 11119--43

\bibitem[{{Cheung} {et~al.}(2015){Cheung}, {Boerner}, {Schrijver}, {Testa},
  {Chen}, {Peter}, \& {Malanushenko}}]{cheung2015}
{Cheung}, M.~C.~M., {Boerner}, P., {Schrijver}, C.~J., {et~al.} 2015, \apj,
  807, 143

\bibitem[{{Christe} {et~al.}(2016){Christe}, {Glesener}, {Buitrago-Casas},
  {Ishikawa}, {Ramsey}, {Gubarev}, {Kilaru}, {Kolodziejczak}, {Watanabe},
  {Takahashi}, {Tajima}, {Turin}, {Shourt}, {Foster}, \&
  {Krucker}}]{Christe2016}
{Christe}, S., {Glesener}, L., {Buitrago-Casas}, C., {et~al.} 2016, Journal of
  Astronomical Instrumentation, 5, 1640005

\bibitem[{{Del Zanna}(2013)}]{zanna2013}
{Del Zanna}, G. 2013, Astronomy and Astrophysics, 558, A73

\bibitem[{{Del Zanna} {et~al.}(2015{\natexlab{a}}){Del Zanna}, {Dere}, {Young},
  {Landi}, \& {Mason}}]{delzanna_etal:2015}
{Del Zanna}, G., {Dere}, K.~P., {Young}, P.~R., {Landi}, E., \& {Mason}, H.~E.
  2015{\natexlab{a}}, \aap, 582, A56

\bibitem[{{Del Zanna} {et~al.}(2015{\natexlab{b}}){Del Zanna}, {Tripathi},
  {Mason}, {Subramanian}, \& {O'Dwyer}}]{Zanna2015}
{Del Zanna}, G., {Tripathi}, D., {Mason}, H., {Subramanian}, S., \& {O'Dwyer},
  B. 2015{\natexlab{b}}, \aap, 573, A104

\bibitem[{{Edlen}(1942)}]{edlen1942}
{Edlen}, B. 1942, Zs.Ap., 22, 30

\bibitem[{Freeland \& Handy(1998)}]{freeland_data_1998}
Freeland, S.~L., \& Handy, B.~N. 1998, Solar Physics, 182, 497

\bibitem[{{Glesener} {et~al.}(2016){Glesener}, {Krucker}, {Christe},
  {Ishikawa}, {Buitrago-Casas}, {Ramsey}, {Gubarev}, {Takahashi}, {Watanabe},
  {Takeda}, {Courtade}, {Turin}, {McBride}, {Shourt}, {Hoberman}, {Foster}, \&
  {Vievering}}]{Glesener2016}
{Glesener}, L., {Krucker}, S., {Christe}, S., {et~al.} 2016, in \procspie, Vol.
  9905, Space Telescopes and Instrumentation 2016: Ultraviolet to Gamma Ray,
  99050E

\bibitem[{{Golub} {et~al.}(2007){Golub}, {Deluca}, {Austin}, {Bookbinder},
  {Caldwell}, {Cheimets}, {Cirtain}, {Cosmo}, {Reid}, {Sette}, {Weber},
  {Sakao}, {Kano}, {Shibasaki}, {Hara}, {Tsuneta}, {Kumagai}, {Tamura},
  {Shimojo}, {McCracken}, {Carpenter}, {Haight}, {Siler}, {Wright}, {Tucker},
  {Rutledge}, {Barbera}, {Peres}, \& {Varisco}}]{golub2007}
{Golub}, L., {Deluca}, E., {Austin}, G., {et~al.} 2007, \solphys, 243, 63

\bibitem[{{Grotrian}(1939)}]{grotrian1939}
{Grotrian}, W. 1939, Naturwissenschaften, 27, 214

\bibitem[{{Gubarev} {et~al.}(2013){Gubarev}, {Ramsey}, {O'Dell}, {Elsner},
  {Kilaru}, {McCracken}, {Pavlinsky}, {Tkachenko}, {Lapshov}, {Atkins}, \&
  {Zavlin}}]{doi:10.1117/12.2027141}
{Gubarev}, M., {Ramsey}, B., {O'Dell}, S.~L., {et~al.} 2013, in \procspie, Vol.
  8861, Optics for EUV, X-Ray, and Gamma-Ray Astronomy VI, 88610K

\bibitem[{{Gubarev} {et~al.}(2014){Gubarev}, {Ramsey}, {Kolodziejczak},
  {O'Dell}, {Elsner}, {Zavlin}, {Swartz}, {Pavlinsky}, {Tkachenko}, \&
  {Lapshov}}]{doi:10.1117/12.2056595}
{Gubarev}, M., {Ramsey}, B., {Kolodziejczak}, J.~J., {et~al.} 2014, in
  \procspie, Vol. 9144, Space Telescopes and Instrumentation 2014: Ultraviolet
  to Gamma Ray, 91444U

\bibitem[{{Hannah} \& {Kontar}(2012)}]{Hannah2012}
{Hannah}, I.~G., \& {Kontar}, E.~P. 2012, \aap, 539, A146

\bibitem[{Hunter(2007)}]{hunter_matplotlib_2007}
Hunter, J.~D. 2007, Computing in Science \& Engineering, 9, 90

\bibitem[{Ishikawa \& Krucker(2019)}]{Ishikawa2019}
Ishikawa, \& Krucker, S. 2019, Hot plasma in a quiescent solar active region as
  measured by RHESSI, XRT, and AIA, , , arXiv:1903.11293

\bibitem[{{Ishikawa} {et~al.}(2011){Ishikawa}, {Saito}, {Tajima}, {Tanaka},
  {Watanabe}, {Odaka}, {Fukuyama}, {Kokubun}, {Takahashi}, {Terada}, {Krucker},
  {Christe}, {McBride}, \& {Glesener}}]{Ishikawa2011}
{Ishikawa}, S., {Saito}, S., {Tajima}, H., {et~al.} 2011, IEEE Transactions on
  Nuclear Science, 58, 2039

\bibitem[{{Ishikawa} {et~al.}(2017){Ishikawa}, {Glesener}, {Krucker},
  {Christe}, {Buitrago-Casas}, {Narukage}, \& {Vievering}}]{Ishikawa2017}
{Ishikawa}, S.-n., {Glesener}, L., {Krucker}, S., {et~al.} 2017, Nature
  Astronomy, 1, 771

\bibitem[{Jones {et~al.}(2001)Jones, Oliphant, \& Peterson}]{jones_scipy_2001}
Jones, E., Oliphant, T., \& Peterson, P. 2001, {{SciPy}}: {{Open}} Source
  Scientific Tools for {{Python}}, ,

\bibitem[{{Klimchuk}(2009)}]{klimchuk2009}
{Klimchuk}, J.~A. 2009, in Astronomical Society of the Pacific Conference
  Series, Vol. 415, Astronomical Society of the Pacific Conference Series, ed.
  {B.~Lites, M.~Cheung, T.~Magara, J.~Mariska, \& K.~Reeves}, 221--+

\bibitem[{{Klimchuk} {et~al.}(2008){Klimchuk}, {Patsourakos}, \&
  {Cargill}}]{klimchuk2008}
{Klimchuk}, J.~A., {Patsourakos}, S., \& {Cargill}, P.~J. 2008, \apj, 682, 1351

\bibitem[{{Kobayashi} {et~al.}(2010){Kobayashi}, {Cirtain}, {Golub}, {Korreck},
  {Cheimets}, {Hertz}, \& {Caldwell}}]{doi:10.1117/12.856793}
{Kobayashi}, K., {Cirtain}, J., {Golub}, L., {et~al.} 2010, in \procspie, Vol.
  7732, Space Telescopes and Instrumentation 2010: Ultraviolet to Gamma Ray,
  773233

\bibitem[{{Kobayashi} {et~al.}(2018){Kobayashi}, {Winebarger}, {Savage},
  {Champey}, {Cheimets}, {Hertz}, {Bruccoleri}, {Scholvin}, {Golub}, {Ramsey},
  {Ranganathan}, {Marquez}, {Allured}, {Parker}, {Heilmann}, \&
  {Schattenburg}}]{doi:10.1117/12.2313997}
{Kobayashi}, K., {Winebarger}, A.~R., {Savage}, S., {et~al.} 2018, in Society
  of Photo-Optical Instrumentation Engineers (SPIE) Conference Series, Vol.
  10699, Space Telescopes and Instrumentation 2018: Ultraviolet to Gamma Ray,
  1069927

\bibitem[{{Kobelski} {et~al.}(2014{\natexlab{a}}){Kobelski}, {McKenzie}, \&
  {Donachie}}]{kobelski2014a}
{Kobelski}, A.~R., {McKenzie}, D.~E., \& {Donachie}, M. 2014{\natexlab{a}},
  \apj, 786, 82

\bibitem[{{Kobelski} {et~al.}(2014{\natexlab{b}}){Kobelski}, {Saar}, {Weber},
  {McKenzie}, \& {Reeves}}]{Kobelski2014}
{Kobelski}, A.~R., {Saar}, S.~H., {Weber}, M.~A., {McKenzie}, D.~E., \&
  {Reeves}, K.~K. 2014{\natexlab{b}}, \solphys, 289, 2781

\bibitem[{{Krivonos} {et~al.}(2017){Krivonos}, {Tkachenko}, {Burenin},
  {Filippova}, {Lapshov}, {Mereminskiy}, {Molkov}, {Pavlinsky}, {Sazonov},
  {Gubarev}, {Kolodziejczak}, {O'Dell}, {Swartz}, {Zavlin}, \&
  {Ramsey}}]{2017ExA....44..147K}
{Krivonos}, R., {Tkachenko}, A., {Burenin}, R., {et~al.} 2017, Experimental
  Astronomy, 44, 147

\bibitem[{{Krucker} {et~al.}(2009{\natexlab{a}}){Krucker}, {Christe},
  {Glesener}, {McBride}, {Turin}, {Glaser}, {Saint-Hilaire}, {Delory}, {Lin},
  {Gubarev}, {Ramsey}, {Terada}, {Ishikawa}, {Kokubun}, {Saito}, {Takahashi},
  {Watanabe}, {Nakazawa}, {Tajima}, {Masuda}, {Minoshima}, \&
  {Shomojo}}]{doi:10.1117/12.827950}
{Krucker}, S., {Christe}, S., {Glesener}, L., {et~al.} 2009{\natexlab{a}}, in
  \procspie, Vol. 7437, Optics for EUV, X-Ray, and Gamma-Ray Astronomy IV,
  743705

\bibitem[{{Krucker} {et~al.}(2009{\natexlab{b}}){Krucker}, {Christe},
  {Glesener}, {McBride}, {Turin}, {Glaser}, {Saint-Hilaire}, {Delory}, {Lin},
  {Gubarev}, {Ramsey}, {Terada}, {Ishikawa}, {Kokubun}, {Saito}, {Takahashi},
  {Watanabe}, {Nakazawa}, {Tajima}, {Masuda}, {Minoshima}, \&
  {Shomojo}}]{krucker2009}
{Krucker}, S., {Christe}, S., {Glesener}, L., {et~al.} 2009{\natexlab{b}}, in
  \procspie, Vol. 7437, Optics for EUV, X-Ray, and Gamma-Ray Astronomy IV,
  743705

\bibitem[{{Krucker} {et~al.}(2011){Krucker}, {Christe}, {Glesener}, {Ishikawa},
  {McBride}, {Glaser}, {Turin}, {Lin}, {Gubarev}, {Ramsey}, {Saito}, {Tanaka},
  {Takahashi}, {Watanabe}, {Tanaka}, {Tajima}, \&
  {Masuda}}]{doi:10.1117/12.895271}
{Krucker}, S., {Christe}, S., {Glesener}, L., {et~al.} 2011, in \procspie, Vol.
  8147, Society of Photo-Optical Instrumentation Engineers (SPIE) Conference
  Series, 814705

\bibitem[{{Krucker} {et~al.}(2013{\natexlab{a}}){Krucker}, {Christe},
  {Glesener}, {Ishikawa}, {Ramsey}, {Gubarev}, {Saito}, {Takahashi},
  {Watanabe}, {Tajima}, {Tanaka}, {Turin}, {Glaser}, {Fermin}, \&
  {Lin}}]{doi:10.1117/12.2024277}
{Krucker}, S., {Christe}, S., {Glesener}, L., {et~al.} 2013{\natexlab{a}}, in
  \procspie, Vol. 8862, Solar Physics and Space Weather Instrumentation V,
  88620R

\bibitem[{{Krucker} {et~al.}(2013{\natexlab{b}}){Krucker}, {Christe},
  {Glesener}, {Ishikawa}, {Ramsey}, {Gubarev}, {Saito}, {Takahashi},
  {Watanabe}, {Tajima}, {Tanaka}, {Turin}, {Glaser}, {Fermin}, \&
  {Lin}}]{krucker2013}
{Krucker}, S., {Christe}, S., {Glesener}, L., {et~al.} 2013{\natexlab{b}}, in
  \procspie, Vol. 8862, Solar Physics and Space Weather Instrumentation V,
  88620R

\bibitem[{{Krucker} {et~al.}(2014){Krucker}, {Christe}, {Glesener}, {Ishikawa},
  {Ramsey}, {Takahashi}, {Watanabe}, {Saito}, {Gubarev}, {Kilaru}, {Tajima},
  {Tanaka}, {Turin}, {McBride}, {Glaser}, {Fermin}, {White}, \&
  {Lin}}]{Krucker2014}
{Krucker}, S., {Christe}, S., {Glesener}, L., {et~al.} 2014, \apjl, 793, L32

\bibitem[{{Lemen} {et~al.}(2012){Lemen}, {Title}, {Akin}, {Boerner}, {Chou},
  {Drake}, {Duncan}, {Edwards}, {Friedlaender}, {Heyman}, {Hurlburt}, {Katz},
  {Kushner}, {Levay}, {Lindgren}, {Mathur}, {McFeaters}, {Mitchell}, {Rehse},
  {Schrijver}, {Springer}, {Stern}, {Tarbell}, {Wuelser}, {Wolfson}, {Yanari},
  {Bookbinder}, {Cheimets}, {Caldwell}, {Deluca}, {Gates}, {Golub}, {Park},
  {Podgorski}, {Bush}, {Scherrer}, {Gummin}, {Smith}, {Auker}, {Jerram},
  {Pool}, {Soufli}, {Windt}, {Beardsley}, {Clapp}, {Lang}, \&
  {Waltham}}]{lemen2012}
{Lemen}, J.~R., {Title}, A.~M., {Akin}, D.~J., {et~al.} 2012, \solphys, 275, 17

\bibitem[{{L{\'o}pez Fuentes} \& {Klimchuk}(2010)}]{lopez2010}
{L{\'o}pez Fuentes}, M.~C., \& {Klimchuk}, J.~A. 2010, \apj, 719, 591

\bibitem[{{O'Dell} {et~al.}(2018){O'Dell}, {Baldini}, {Bellazzini}, {Costa},
  {Elsner}, {Kaspi}, {Kolodziejczak}, {Latronico}, {Marshall}, {Matt},
  {Mulieri}, {Ramsey}, {Romani}, {Soffitta}, {Tennant}, {Weisskopf}, {Allen},
  {Amici}, {Antoniak}, {Attina}, {Bachetti}, {Barbanera}, {Baumgartner},
  {Bladt}, {Bongiorno}, {Borotto}, {Brooks}, {Bussinger}, {Bygott},
  {Cavazzuti}, {Ceccanti}, {Citraro}, {Deininger}, {Del Monte}, {Dietz}, {Di
  Lalla}, {Di Persio}, {Donnarumma}, {Erickson}, {Evangelista}, {Fabiani},
  {Ferrazzoli}, {Foster}, {Giusti}, {Gunji}, {Guy}, {Johnson}, {Kalinowski},
  {Kelley}, {Kilaru}, {Lefevre}, {Maldera}, {Manfreda}, {Marengo},
  {Masciarelli}, {McEachen}, {Mereu}, {Minuti}, {Mitchell}, {Mitchell},
  {Mitsuishi}, {Morbidini}, {Mosti}, {Nasimi}, {Negri}, {Orsini}, {Osborne},
  {Pavelitz}, {Pentz}, {Perri}, {Pesce-Rollins}, {Peterson}, {Piazzolla},
  {Pieraccini}, {Pilia}, {Pinchera}, {Puccetti}, {Ranganathan}, {Read},
  {Rubini}, {Santoli}, {Sarra}, {Schindhelm}, {Sciortino}, {Seckar},
  {Sgr{\`o}}, {Smith}, {Speegle}, {Tamagawa}, {Tardiola}, {Tobia}, {Tortosa},
  {Trois}, {Weddendorf}, {Wedmore}, \& {Zanetti}}]{stephen2018}
{O'Dell}, S.~L., {Baldini}, L., {Bellazzini}, R., {et~al.} 2018, in Society of
  Photo-Optical Instrumentation Engineers (SPIE) Conference Series, Vol. 10699,
  Space Telescopes and Instrumentation 2018: Ultraviolet to Gamma Ray, 106991X

\bibitem[{{O'Dwyer} {et~al.}(2010){O'Dwyer}, {Del Zanna}, {Mason}, {Weber}, \&
  {Tripathi}}]{odwyer2010}
{O'Dwyer}, B., {Del Zanna}, G., {Mason}, H.~E., {Weber}, M.~A., \& {Tripathi},
  D. 2010, \aap, 521, A21

\bibitem[{Oliphant(2006)}]{oliphant_guide_2006}
Oliphant, T. 2006, A {{Guide}} to {{Numpy}} ({USA}: {Trelgol Publishing})

\bibitem[{{Parker}(1983{\natexlab{a}})}]{parker1983b}
{Parker}, E.~N. 1983{\natexlab{a}}, \apj, 264, 642

\bibitem[{{Parker}(1983{\natexlab{b}})}]{parker1983a}
---. 1983{\natexlab{b}}, \apj, 264, 635

\bibitem[{P{\'e}rez \& Granger(2007)}]{perez_ipython_2007}
P{\'e}rez, F., \& Granger, B.~E. 2007, Computing in Science \& Engineering, 9,
  21

\bibitem[{Pesnell {et~al.}(2012)Pesnell, Thompson, \&
  Chamberlin}]{pesnell_solar_2012}
Pesnell, W.~D., Thompson, B.~J., \& Chamberlin, P.~C. 2012, Solar Physics, 275,
  3

\bibitem[{{Reale} {et~al.}(2009){Reale}, {Testa}, {Klimchuk}, \&
  {Parenti}}]{reale2009}
{Reale}, F., {Testa}, P., {Klimchuk}, J.~A., \& {Parenti}, S. 2009, \apj, 698,
  756

\bibitem[{{Reep} {et~al.}(2013){Reep}, {Bradshaw}, \& {Klimchuk}}]{reep2013a}
{Reep}, J.~W., {Bradshaw}, S.~J., \& {Klimchuk}, J.~A. 2013, \apj, 764, 193

\bibitem[{{Schmelz} {et~al.}(2012){Schmelz}, {Reames}, {von Steiger}, \&
  {Basu}}]{schmelz2012}
{Schmelz}, J.~T., {Reames}, D.~V., {von Steiger}, R., \& {Basu}, S. 2012, \apj,
  755, 33

\bibitem[{{Schmelz} {et~al.}(2009{\natexlab{a}}){Schmelz}, {Saar}, {DeLuca},
  {Golub}, {Kashyap}, {Weber}, \& {Klimchuk}}]{schmelz2009a}
{Schmelz}, J.~T., {Saar}, S.~H., {DeLuca}, E.~E., {et~al.} 2009{\natexlab{a}},
  \apjl, 693, L131

\bibitem[{{Schmelz} {et~al.}(2009{\natexlab{b}}){Schmelz}, {Kashyap}, {Saar},
  {Dennis}, {Grigis}, {Lin}, {De Luca}, {Holman}, {Golub}, \&
  {Weber}}]{schmelz2009b}
{Schmelz}, J.~T., {Kashyap}, V.~L., {Saar}, S.~H., {et~al.} 2009{\natexlab{b}},
  \apj, 704, 863

\bibitem[{{Shestov} {et~al.}(2010){Shestov}, {Kuzin}, {Urnov}, {Ul'Yanov}, \&
  {Bogachev}}]{shestov2010}
{Shestov}, S.~V., {Kuzin}, S.~V., {Urnov}, A.~M., {Ul'Yanov}, A.~S., \&
  {Bogachev}, S.~A. 2010, Astronomy Letters, 36, 44

\bibitem[{{Teriaca} {et~al.}(2012){Teriaca}, {Warren}, \&
  {Curdt}}]{teriaca2012}
{Teriaca}, L., {Warren}, H.~P., \& {Curdt}, W. 2012, \apjl, 754, L40

\bibitem[{{Testa} {et~al.}(2012){Testa}, {De Pontieu},
  {Mart{\'{\i}}nez-Sykora}, {Hansteen}, \& {Carlsson}}]{testa2012}
{Testa}, P., {De Pontieu}, B., {Mart{\'{\i}}nez-Sykora}, J., {Hansteen}, V., \&
  {Carlsson}, M. 2012, \apj, 758, 54

\bibitem[{{Testa} {et~al.}(2011){Testa}, {Reale}, {Landi}, {DeLuca}, \&
  {Kashyap}}]{Testa11}
{Testa}, P., {Reale}, F., {Landi}, E., {DeLuca}, E.~E., \& {Kashyap}, V. 2011,
  \apj, 728, 30

\bibitem[{{The Astropy Collaboration} {et~al.}(2018){The Astropy
  Collaboration}, {Price-Whelan}, Sip{\H o}cz, G{\"u}nther, Lim, Crawford,
  Conseil, Shupe, Craig, Dencheva, Ginsburg, VanderPlas, Bradley,
  {P{\'e}rez-Su{\'a}rez}, {de Val-Borro}, Paper~Contributors, Aldcroft, Cruz,
  Robitaille, Tollerud, Coordination~Committee, Ardelean, Babej, Bach,
  Bachetti, Bakanov, Bamford, Barentsen, Barmby, Baumbach, Berry, Biscani,
  Boquien, Bostroem, Bouma, Brammer, Bray, Breytenbach, Buddelmeijer, Burke,
  Calderone, Cano~Rodr{\'i}guez, Cara, Cardoso, Cheedella, Copin, Corrales,
  Crichton, D'Avella, Deil, Depagne, Dietrich, Donath, Droettboom, Earl, Erben,
  Fabbro, Ferreira, Finethy, Fox, Garrison, Gibbons, Goldstein, Gommers, Greco,
  Greenfield, Groener, Grollier, Hagen, Hirst, Homeier, Horton, Hosseinzadeh,
  Hu, Hunkeler, Ivezi{\'c}, Jain, Jenness, Kanarek, Kendrew, Kern, Kerzendorf,
  Khvalko, King, Kirkby, Kulkarni, Kumar, Lee, Lenz, Littlefair, Ma, Macleod,
  Mastropietro, McCully, Montagnac, Morris, Mueller, Mumford, Muna, Murphy,
  Nelson, Nguyen, Ninan, N{\"o}the, Ogaz, Oh, Parejko, Parley, Pascual, Patil,
  Patil, Plunkett, Prochaska, Rastogi, Reddy~Janga, Sabater, Sakurikar,
  Seifert, Sherbert, {Sherwood-Taylor}, Shih, Sick, Silbiger, Singanamalla,
  Singer, Sladen, Sooley, Sornarajah, Streicher, Teuben, Thomas, Tremblay,
  Turner, Terr{\'o}n, {van Kerkwijk}, {de la Vega}, Watkins, Weaver, Whitmore,
  Woillez, Zabalza, \& Contributors}]{the_astropy_collaboration_astropy_2018}
{The Astropy Collaboration}, {Price-Whelan}, A.~M., Sip{\H o}cz, B.~M.,
  {et~al.} 2018, The Astronomical Journal, 156, 123

\bibitem[{{Tripathi} {et~al.}(2011){Tripathi}, {Klimchuk}, \&
  {Mason}}]{tripathi2011}
{Tripathi}, D., {Klimchuk}, J.~A., \& {Mason}, H.~E. 2011, \apj, 740, 111

\bibitem[{{Tripathi} {et~al.}(2010){Tripathi}, {Mason}, \&
  {Klimchuk}}]{tripathi2010a}
{Tripathi}, D., {Mason}, H.~E., \& {Klimchuk}, J.~A. 2010, \apj, 723, 713

\bibitem[{{van Ballegooijen} {et~al.}(2014){van Ballegooijen}, {Asgari-Targhi},
  \& {Berger}}]{vanballegooijen2014}
{van Ballegooijen}, A.~A., {Asgari-Targhi}, M., \& {Berger}, M.~A. 2014, \apj,
  787, 87

\bibitem[{{van Ballegooijen} {et~al.}(2011){van Ballegooijen}, {Asgari-Targhi},
  {Cranmer}, \& {DeLuca}}]{vanballegooijen2011}
{van Ballegooijen}, A.~A., {Asgari-Targhi}, M., {Cranmer}, S.~R., \& {DeLuca},
  E.~E. 2011, \apj, 736, 3

\bibitem[{{Viall} \& {Klimchuk}(2011)}]{viall2011}
{Viall}, N.~M., \& {Klimchuk}, J.~A. 2011, \apj, 738, 24

\bibitem[{{Viall} \& {Klimchuk}(2012)}]{viall2012}
---. 2012, \apj, 753, 35

\bibitem[{{Viall} \& {Klimchuk}(2013)}]{viall2013}
---. 2013, \apj, 771, 115

\bibitem[{{Warren} {et~al.}(2012){Warren}, {Winebarger}, \&
  {Brooks}}]{warren2012}
{Warren}, H.~P., {Winebarger}, A.~R., \& {Brooks}, D.~H. 2012, \apj, 759, 141

\bibitem[{Waskom {et~al.}(2018)Waskom, Botvinnik, O'Kane, Hobson, Ostblom,
  Lukauskas, Gemperline, Augspurger, Halchenko, Cole, Warmenhoven, Ruiter, Pye,
  Hoyer, Vanderplas, Villalba, Kunter, Quintero, Bachant, Martin, Meyer, Miles,
  Ram, Brunner, Yarkoni, Williams, Evans, Fitzgerald, ~, \&
  Qalieh}]{waskom_seaborn_2018}
Waskom, M., Botvinnik, O., O'Kane, D., {et~al.} 2018, Seaborn: V0.9.0,
  {Zenodo}, doi:10.5281/zenodo.1313201

\bibitem[{{Winebarger} {et~al.}(2011){Winebarger}, {Schmelz}, {Warren}, {Saar},
  \& {Kashyap}}]{winebarger2011}
{Winebarger}, A.~R., {Schmelz}, J.~T., {Warren}, H.~P., {Saar}, S.~H., \&
  {Kashyap}, V.~L. 2011, \apj, 740, 2

\bibitem[{{Winebarger} {et~al.}(2012){Winebarger}, {Warren}, {Schmelz},
  {Cirtain}, {Mulu-Moore}, {Golub}, \& {Kobayashi}}]{winebarger2012}
{Winebarger}, A.~R., {Warren}, H.~P., {Schmelz}, J.~T., {et~al.} 2012, \apjl,
  746, L17

\end{thebibliography}
\end{document}